\documentclass[lettersize,journal]{IEEEtran}
\usepackage{makecell}
\usepackage{subfigure}
\usepackage{amsmath}
\usepackage{amsmath}
\usepackage{multirow}
\usepackage{cite}
\usepackage{amsmath,amssymb,amsfonts}
\usepackage{algorithmic}
\usepackage{graphicx}
\usepackage{textcomp}
\usepackage{xcolor}
\usepackage{stfloats}
\usepackage{booktabs}
\usepackage{amsmath,amsfonts}
\usepackage{algorithmic}
\usepackage{array}
\usepackage[caption=false,font=normalsize,labelfont=sf,textfont=sf]{subfig}
\usepackage{textcomp}
\usepackage{stfloats}
\usepackage{url}
\usepackage{verbatim}
\usepackage{graphicx}
\usepackage[ruled,linesnumbered]{algorithm2e}
\def\BibTeX{{\rm B\kern-.05em{\sc i\kern-.025em b}\kern-.08em
T\kern-.1667em\lower.7ex\hbox{E}\kern-.125emX}}
\usepackage{balance}
\begin{document}

\title{
Cooperative Mapping, Localization, and Beam Management via Multi-Modal SLAM in ISAC Systems
}

\author{
Hang~Que,~\IEEEmembership{Graduate~Student~Member,~IEEE,}
Jie~Yang,~\IEEEmembership{Member,~IEEE,}
Tao~Du,~\IEEEmembership{Graduate~Student~Member,~IEEE,}
Shuqiang~Xia,~
Chao-Kai~Wen,~\IEEEmembership{Fellow,~IEEE,}
and~Shi~Jin,~\IEEEmembership{Fellow,~IEEE}

\thanks{
Hang Que, Tao Du, and Shi Jin are with the National Mobile Communications Research Laboratory, Southeast University, Nanjing 211189, China (e-mail: \{quehang; dutao; jinshi\}@seu.edu.cn).

Jie Yang is with the Key Laboratory of Measurement and Control of Complex Systems of Engineering, Ministry of Education, Southeast University, Nanjing 210096, China (e-mail: yangjie@seu.edu.cn).

Jie Yang and Shi Jin are with the Frontiers Science Center for Mobile Information Communication and Security, Southeast University, Nanjing 210096, China.
Chao-Kai Wen is with the Institute of Communications Engineering, National Sun Yat-sen University, Kaohsiung 80424, Taiwan. (e-mail: chaokai.wen@mail.nsysu.edu.tw).

Shuqiang Xia is with the ZTE Corporation and the State Key Laboratory of Mobile Network and Mobile Multimedia, Shenzhen 518055, China (e-mail: xia.shuqiang@zte.com.cn).
}
}

\markboth{IEEE TRANSACTIONS,~Vol.~XX, No.~XX,~XX~20XX}%
{Shell \MakeLowercase{\textit{et al.}}: Cooperative Mapping, Localization, and Beam Management in Multi-Modal ISAC Systems}

\maketitle

\begin{abstract}

Simultaneous localization and mapping (SLAM) plays a critical role in integrated sensing and communication (ISAC) systems for sixth-generation (6G) millimeter-wave (mmWave) networks, enabling environmental awareness and precise user equipment (UE) positioning. 
While cooperative multi-user SLAM has demonstrated potential in leveraging distributed sensing, its application within multi-modal ISAC systems remains limited, particularly in terms of theoretical modeling and communication-layer integration. 
This paper proposes a novel multi-modal SLAM framework that addresses these limitations through three key contributions. First, a Bayesian estimation framework is developed for cooperative multi-user SLAM, along with a two-stage algorithm for robust radio map construction under dynamic and heterogeneous sensing conditions. 
Second, a multi-modal localization strategy is introduced, fusing SLAM results with camera-based multi-object tracking and inertial measurement unit (IMU) data via an error-aware model, significantly improving UE localization in multi-user scenarios. 
Third, a sensing-aided beam management scheme is proposed, utilizing global radio maps and localization data to generate UE-specific prior information for beam selection, thereby reducing inter-user interference and enhancing downlink spectral efficiency. Simulation results demonstrate that the proposed system improves radio map accuracy by up to 60\%, enhances localization accuracy by 37.5\%, and significantly outperforms traditional methods in both indoor and outdoor environments.  
\end{abstract}
\vspace{0.2cm}

\section{Introduction}
\IEEEPARstart{I}{ntegrated} sensing and communication (ISAC) has emerged as a key enabling technology for sixth-generation (6G) millimeter-wave (mmWave) wireless systems, offering simultaneous high-quality communication and advanced sensing services. These services include high-precision positioning, human gesture recognition, and environmental mapping, and have attracted significant attention from both academia and industry \cite{b0, b1, b2}. Among various ISAC innovations, simultaneous localization and mapping (SLAM) stands out as a foundational framework for 6G mmWave systems, providing advanced sensing capabilities \cite{b3, b4} and supporting sensing-assisted communication \cite{b5,b6}.

SLAM primarily focuses on the concurrent localization of user equipment (UE) and the construction of environmental or radio maps, which capture key features necessary for modeling mmWave multipath propagation \cite{a201, a6, a3, a2, a4, a5}. The development of SLAM algorithms often involves Bayesian frameworks to estimate spatial parameters from uncertain observations. Existing methodologies span geometry-based techniques \cite{a201, a6}, random finite set approaches \cite{a3}, and belief propagation-based algorithms \cite{a2, a4}. However, the reliance on UE-side execution of SLAM poses challenges due to UE mobility and limited sensing capabilities. These constraints often result in inaccuracies and frequent updates to radio maps, underscoring the need for more robust SLAM strategies in dynamic mmWave environments.

To overcome these limitations, cooperative multi-user SLAM has emerged as a promising solution. In this paradigm, multiple UEs collaborate to perform SLAM and upload their results to the base station (BS), which fuses these data into a global radio map. UEs can then utilize this map to refine their own SLAM estimations. However, due to varying fields of view (FoVs), mobility, and sensing capabilities, UEs often generate local maps with minimal overlap, making fusion difficult. Additionally, asynchronous sensing and access among UEs introduce further design complexities. 
Several studies have attempted to address individual aspects of these challenges. For instance, crowdsourcing-inspired approaches allow UEs to upload only their most reliable data \cite{a5}. The authors in \cite{R1C201} investigated cooperative multi-user localization with a known radio map, while \cite{R1C202} explored joint localization and mapping based on multi-user time delay measurements. Moreover, \cite{R1C203} explores cooperative construction of a scatterer point cloud map, integrating it with camera images through deep learning techniques. Other efforts focus on data association using Mahalanobis distance and fusion techniques like arithmetic averaging (AA) and generalized covariance intersection (GCI) \cite{a6}. In addition, shared virtual transmitters and reflective surfaces have been modeled to support cooperative channel SLAM \cite{a7}.

Despite these advances, two major gaps persist. First, existing works typically address isolated components of the multi-user SLAM pipeline, lacking a comprehensive theoretical estimation framework that models the entire process. While some algorithms achieve good performance in specific scenarios, they often lack scalability and generalizability. A unified theoretical foundation is essential for developing robust, context-aware solutions. Second, current literature predominantly leverages SLAM-generated radio maps for localization but neglects the broader potential of SLAM-derived sensing data to enhance mmWave communications, particularly in beam management.


Beamforming is critical in mmWave communication but incurs significant overhead for beam management \cite{a8, a9}. Recent studies have sought to reduce this overhead by utilizing prior beam measurements, UE mobility, and environmental information \cite{a10, a11, a12, a13}. Multi-modal ISAC systems equipped with diverse sensors, such as cameras and inertial measurement units (IMUs), offer promising directions for beam optimization \cite{a14, a1401, a1402, a1403, a19}. In particular, fixed cameras can detect UE motion, and some works demonstrate that BS-mounted cameras can predict line-of-sight (LoS) beam directions \cite{a15, a16, a19, a20}.

In multi-user scenarios, managing inter-user interference (IUI) becomes a key challenge, especially when traditional sensors struggle to distinguish mmWave UEs from other environmental objects \cite{a21, a22, a23,a2301}. While multi-user matching techniques in multi-modal systems offer partial solutions \cite{a24, a25}, effective beam management in dynamic environments remains elusive. One promising avenue is to combine radio SLAM with sensor data to enable accurate multi-user identification and association. Although Visual SLAM and LiDAR SLAM have demonstrated successful multi-sensor fusion \cite{R1C504, R1C505}, the integration of diverse sensing modalities into radio SLAM remains underexplored. Inadequate fusion can impair sensing accuracy and exacerbate interference, especially in dense multi-user deployments, highlighting the need for more effective multi-modal SLAM solutions.


In response to the aforementioned challenges, this work proposes a unified multi-modal SLAM framework that enhances radio map construction, UE localization, and multi-user beam management for ISAC systems. The key contributions are as follows:

\begin{itemize}
\item \textbf{Cooperative Radio Map Construction:}  
We develop a novel Bayesian framework tailored for cooperative multi-user SLAM, modeling key processes as probabilistic inference tasks. A two-stage SLAM algorithm is proposed: the initialization stage processes local UE results to build an initial radio map in the absence of a reliable global map; the refinement stage further improves the map using stable global information, enhancing robustness and adaptability. 

\item \textbf{Multi-Modal Enhanced Localization:}  
A hybrid localization scheme is introduced, integrating SLAM, fixed stereo camera data, and IMU measurements. The stereo camera employs the YOLO algorithm for multi-object detection and localization. An offline-trained error model fuses camera-based localization with SLAM outputs, supported by IMU data, to enhance UE localization in multi-modal contexts.

\item \textbf{Sensing-Aided Beam Management:}  
A novel beam management strategy is proposed, which utilizes the global radio map and multi-user localization data derived from SLAM to optimize communication. This method reduces IUI by constructing informed priors on channel path distributions, which are then allocated to UEs for optimized beam alignment. This process enhances the downlink spectral efficiency (SE) in multi-user ISAC scenarios.

\end{itemize}

The remainder of this study is organized as follows: Section II describes the system model; Sections III and IV present the multi-user SLAM Bayesian framework and its implementation; Section V details the multi-modal enhanced localization algorithm; Section VI explains the sensing-aided beam management algorithm; Section VII demonstrates simulation results under various settings; and Section VIII concludes the study.

\textbf{Notations---}In this paper, the notation
${\mathbf A}$ is a matrix, ${\mathbf a}$ is a vector, $a$ is a scalar, and ${\mathbb A}$ is a set.
$( \cdot )^{\text{T}}$, $( \cdot )^{\text{H}}$, and $( \cdot )^{\text{-1}}$ denote the transpose, conjugate transpose, and matrix inversion operation, respectively.
The cardinality of set ${\mathbb A}$ is denoted by $|{{\mathbb A}}|$.
The ${l_2}$ norm of vector ${\mathbf a}$ is denoted by $\|{{\mathbf a}}\|_2$.
For scalar ${ a}$, $|{a}|$ indicates its absolute value, and $\lceil{a}\rceil$ represents the smallest integer greater than or equal to ${ a}$.
The notation ${\mathcal{N}}( {\mu,\sigma^2} )$ denotes a Gaussian distribution with mean $\mu$ and variance $\sigma^2$.
The notation $\mathcal{N}_d( \boldsymbol{\mu}, {\bf{\Sigma}} )$ 
denotes a $d$-dimensional multivariate Gaussian distribution with mean vector $\boldsymbol{\mu} \in \mathbb{R}^d$ and covariance matrix ${\bf{\Sigma}} \in \mathbb{R}^{d \times d}$. The notation ${\text{diag}}(a,b)$ denotes a $2 \times 2$ diagonal matrix with entries $a$ and $b$ on its main diagonal. 
The indicator function $1_{(a > b)}$ equals 1 if $a > b$, and 0 otherwise.

\vspace{-0.2cm}
\section{System Model}\label{II}
We consider a mmWave multi-modal ISAC system consisting of a static BS and multiple moving UEs.
Both the BS and UEs are equipped with multiple antenna elements, enabling beamforming for communication.
In addition, each UE is equipped with a channel azimuth estimator and an IMU, while the BS is equipped with a static stereo camera.
In this paper, the channel estimator, IMU, and camera are collectively referred to as sensors, which can provide measurements for implementing SLAM.
In this section, we first introduce the channel and signal models for the communication system. Then, we describe the models for the sensor measurements.

\begin{figure}
\centering
\includegraphics[scale=0.35]{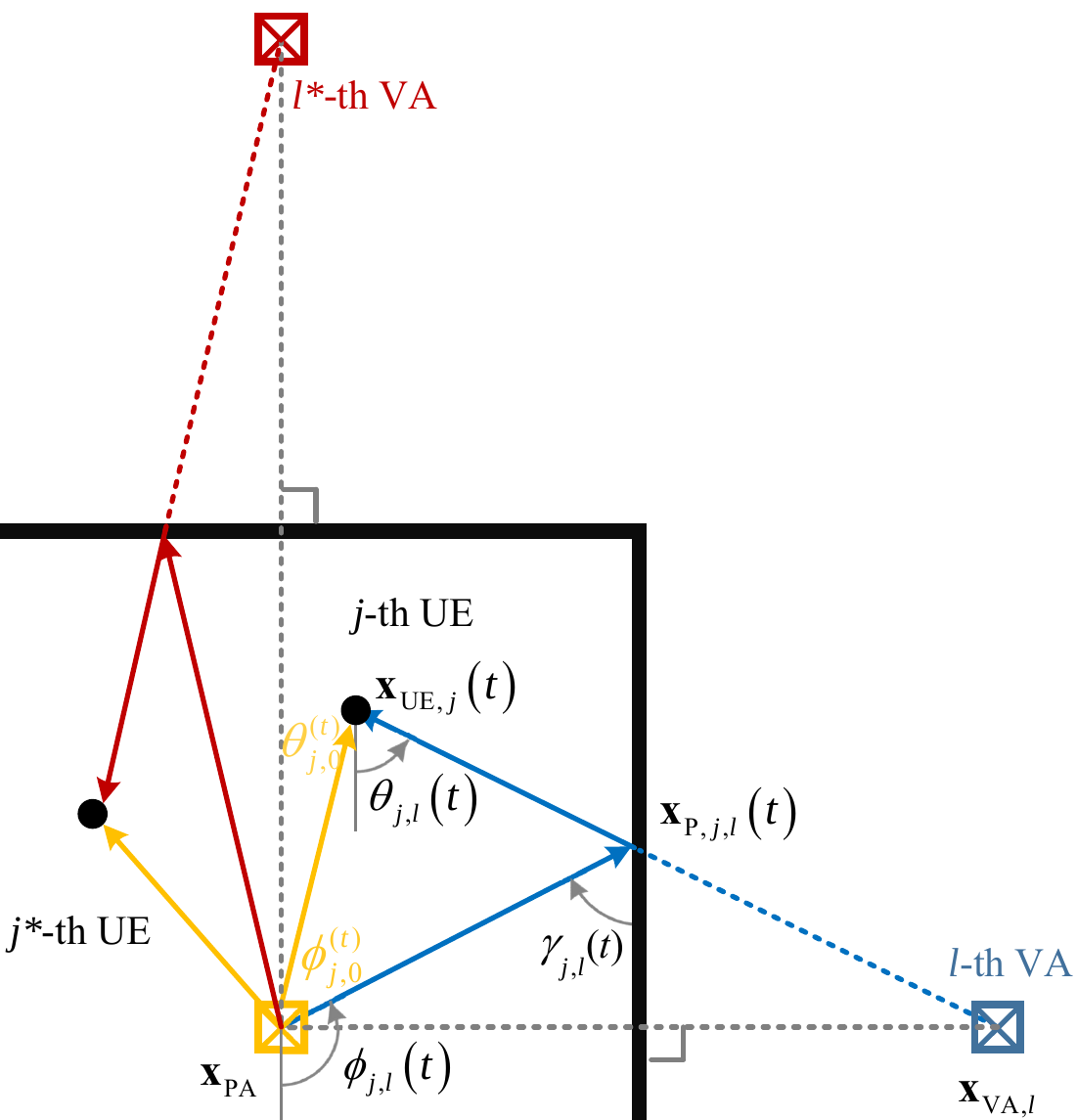}
\caption{An illustration of the mmWave channel multipath propagation process.}
\label{f1}
\vspace{-0.2cm}
\end{figure}

\subsection{Signal and Channel Models}

The BS equipped with $N_{\text{BS}}$ antenna elements serves a total of $J$ UEs, each featuring $N_{\text{UE}}$ antenna elements.
The BS communicates with each UE using a single radio-frequency chain to implement beamforming.
The received signal at the $j$-th UE, denoted as ${r_j}(t)$, is given by
\begin{equation}
{r_j}(t)={{\bf{w}}^{\text{H}}_{j}}(t){{\bf{H}}_j}(t) {\bf{f}}_{j}(t) +{{\bf{w}}^{\text{H}}_{j}}(t){\bf n},
\end{equation}
where ${\bf n}\sim \mathcal{N}_{N_{\text{UE}}}( {{\bf 0},\sigma ^2 {\bf I}} )$ represents the additive Gaussian noise vector. The term $\mathbf{H}_{j}{(t)}\in {\mathbb{C}^{{N_{{\text{UE}}}} \times{N_{{\text{BS}}}}}}$ specifies the downlink channel matrix between the BS and the $j$-th UE at time $t$, and the vectors ${\bf{f}}_{j}( t )\in {\mathbb{C}^{ {N_{{\text{BS}}}} \times{1} } }$ and ${\bf{w}}_{j}( t )\in {\mathbb{C}^{ {N_{{\text{UE}}}} \times{1} } }$ denote the beamforming and combining vectors, respectively.

In our model, the mmWave channel from the BS to the $j$-th UE $\mathbf{H}_{j}{(t)}\in {\mathbb{C}^{{N_{{\text{UE}}}} \times{N_{{\text{BS}}}}}}$ is depicted as follows:
\begin{equation}
{{\bf{H}}_j}( t ) = \mathop \sum \limits_{l = 1}^{{L_j}(t)} {g_{j,l}(t)}{{\bf{a}}_{{\text{UE}}}}{{\left( {{\theta _{j,l}}(t)} \right)}}{\bf{a}}_{{\text{BS}}}^{\text{H}}{{\left( {{\phi _{j,l}}(t)} \right)}},
\label{e1}
\end{equation}
where ${{\bf{a}}_{{\text{UE}}}}( \cdot)$ and ${{\bf{a}}_{{\text{BS}}}}( \cdot)$ represent the steering vectors of the UE and BS antenna arrays, respectively. The channel parameters ${g_{j,l}}(t)$, ${{\theta _{j,l}}( t )}$, and ${{\phi _{j,l}}( t )}$ are the complex gain, angle of arrival (AoA), and angle of departure (AoD) of the $l$-th path, respectively, with $L_j(t)$ indicating the total number of available paths including the LoS and specular single-bounce NLoS paths.\footnote{NLoS paths undergoing two or more reflections are assumed to have negligible path gain due to substantial attenuation, thus modeled as part of the Gaussian noise.}

The propagation of the LoS path is visualized in yellow in Fig. \ref{f1}.
The position of the $j$-th UE at time $t$ is defined as ${{\bf{x}}_{{\text{UE}},j}}( t ) = [ {{x_{{\text{UEX}},j}}( t ),{x_{{\text{UEY}},j}}( t )} ]$, and the BS position as ${{\bf{x}}_{{\text{PA}}}} = [ {{x_{{\text{PAX}}}},{x_{{\text{PAY}}}}} ]$. The azimuth angles ${\theta _{j,l}^{(t)}}$, ${\phi _{j,l}^{(t)}}$ for the LoS path and the path gain $g_{j,l}(t)$ are computed as follows
\begin{subequations}
\label{e3}
\begin{align}
{\tan}{\theta _{j,l}}(t) &={\tan}{\phi _{j,l}}(t) = \frac{{{x_{{\text{UEY}},j}}(t) - {x_{\text{PAY}}}{\text{\;}}}}{{{x_{{\text{UEX}},j}}(t) - {x_{\text{PAX}}}{\text{\;}}}},\\
{g}_{j,l}(t) &\propto \frac{1}{\left\| {{\bf{x}}_{\text{UE},j}(t) - {\bf{x}}_{\text{PA}}} \right\|_2},\label{e3b}
\end{align}
\end{subequations}
where the path loss model accounts for propagation loss \cite{PathGain}.

NLoS path propagation, depicted in blue in Fig. \ref{f1}, involves reflection off a wall with the reflection point ${{\bf{x}}_{{\text{P}},j,l}}( t ) = [ {{x_{{\text{PX}},j,l}}( t ),{x_{{\text{PY}},j,l}}( t )} ]$. The mirror image of the BS, forming the $l$-th NLoS path, is identified as the $l$-th virtual anchor (VA) located at ${{\bf{x}}_{{\text{VA}},l}} = [ {{x_{{\text{VAX}},l}},{x_{{\text{VAY}},l}}} ]$.
The azimuth angles ${{\theta _{j,l}}( t )}$, ${{\phi _{j,l}}( t )}$ for the NLoS path and the path gain $g_{j,l}(t)$ are given by
\begin{subequations}
\label{e4}
\begin{align}
{\tan}{\theta _{j,l}}(t) &= \frac{{{x_{{\text{UEY}},j}}(t) - {x_{{\text{VAY}},l}}{\text{\;}}}}{{{x_{{\text{UEX}},j}}(t) - {x_{{\text{VAX}},l}}{\text{\;}}}},\\
{\tan}{\phi _{j,l}}(t) &= \frac{{{x_{{\text{PY}},j,l}}(t) - {x_{{\text{PAY}}}}{\text{\;}}}}{{{x_{{\text{PX}},j,l}}(t) - {x_{{\text{PAX}}}}{\text{\;}}}},\\
{g}_{j,l}(t) &\propto \frac{{\left| \frac{{\sin{\gamma_{j,l}}(t) - \sqrt{\varepsilon - \cos^2{\gamma_{j,l}}(t)}}}{{\sin{\gamma_{j,l}}(t) + \sqrt{\varepsilon - \cos^2{\gamma_{j,l}}(t)}}} \right|} }{{\left\| {{\bf{x}}_{\text{UE},j}(t) - {\bf{x}}_{\text{VA},l}} \right\|}_2},\label{e4c}
\end{align}
\end{subequations}
with $\gamma_{j,l}(t)=(\pi-|{\theta _{j,l}}( t )-{\phi _{j,l}}( t )|)/2$ being the reflection angle and $\varepsilon$ the dielectric constant of the wall. The model includes both propagation and reflection losses \cite{PathGain}.
Identifying the BS as the PA and the VAs as features, these elements constitute a radio map delineating the LoS and NLoS propagation paths. The SLAM challenge involves concurrently achieving accurate UE localization and constructing this radio map.

\begin{figure*}
\centering
\includegraphics[scale=0.75]{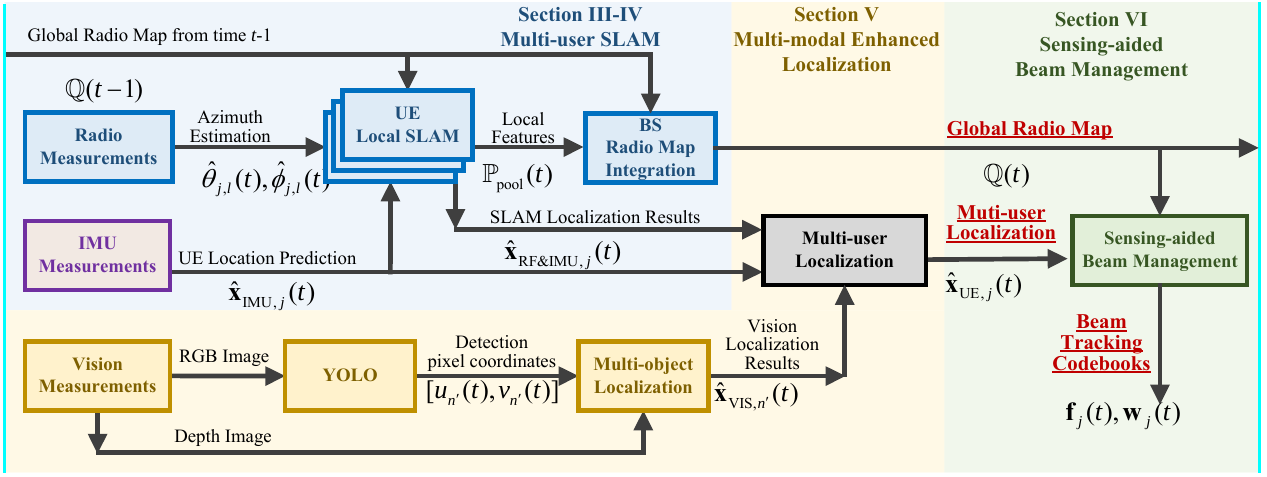}
\caption{Block diagram of the proposed cooperative radio map construction, UE localization, and beam management algorithm.}\label{f2}\vspace{-2mm}
\end{figure*}

The beamforming vectors ${\bf{f}}_{j}( t )$ and ${\bf{w}}_{j}( t )$, utilized at the BS and the UE sides, respectively, are chosen from predefined codebooks \cite{a8}, adhering to the following constraints
\begin{subequations}
\label{e5}
\begin{align}
{\bf{f}}_{j}(t) \in& \left\{ {{\boldsymbol{\sf{f}}_{m}}{\Big{|}}m = 1, \ldots ,M} \right\},\\
{{\bf{w}}_{j}(t)} \in& \left\{ {{\boldsymbol{\sf{w}}_{m}}{\Big{|}}m = 1, \ldots ,M} \right\},
\end{align}
\end{subequations}
where the beamforming vectors ${{\boldsymbol{\sf{f}}_{m}}\in {\mathbb{C}^{ {N_{{\text{BS}}}} \times{1} } }}$ and ${\boldsymbol{\sf{w}}_{m}}\in {\mathbb{C}^{ {N_{{\text{UE}}}} \times{1} } }$ are oriented in the $\frac{m}{M}\pi$ direction, featuring a codebook angular resolution of $\pi/M$.
Furthermore, the SE of all UEs can be computed as
\begin{equation}\label{e6}
{\text{SE}}{\left(t\right)} = \mathop \sum \limits_{j = 1}^J {{\text{log}}_2}{{\left( {1 + \frac{{{{{\left| {{\mathbf{w}}_j^{\text{H}}(t){{\mathbf{H}}_j}(t){{\mathbf{f}}_j}(t)} \right|}}^2}}}{{\mathop \sum \nolimits_{i \ne j} {{{\left| {{\mathbf{w}}_j^{\text{H}}(t){{\mathbf{H}}_j}(t){{\mathbf{f}}_i}(t)} \right|}}^2} + {\sigma ^2}}}} \right)}}.
\end{equation}
The challenge of multi-user beam management is thus to select the most suitable ${\bf{f}}_{j}( t )$ and ${\bf{w}}_{j}( t )$ to maximize ${\text{SE}}(t)$.

\subsection{Measurement Models}
To facilitate the SLAM function and enhance beam management capabilities, various sensors are deployed at both the BS and UE sides. The functionality and measurements derived from these sensors are as follows:

Each UE is equipped with a channel estimator and an IMU. The channel estimator is capable of obtaining radio measurements from the mmWave channel, as described in (\ref{e1}), yielding estimates of the AoA and AoD, denoted as ${{\hat{\theta} _{j,l}}( t )}$ and ${{\hat{\phi} _{j,l}}( t )}$, respectively. Notably, the channel estimator model does not include time-of-arrival measurements due to the significant challenge of precisely synchronizing clocks across multiple UEs. These estimates are modeled as outcomes of Gaussian random variables:
\begin{equation}
\label{e7}
{\left( {{\hat \theta }_{j,l}}(t), {{\hat \phi }_{j,l}}(t) \right)} \sim {\left( \mathcal{N}( {{\theta _{j,l}}(t),\sigma _{{\text{angle}}}^2} ) , \mathcal{N}( {{\phi _{j,l}}(t),\sigma _{{\text{angle}}}^2} ) \right)}
\end{equation}
for $l = 1,\ldots ,{{\hat L}_j}(t)$, where ${{\hat L}_j}(t)$ represents the estimated number of channel paths.
The IMU provides relative location and orientation data, with the predicted location of ${{\bf{x}}_{{\text{UE}},j}}( t )$, incorporating acceleration noise variance by the IMU denoted as $\sigma _{{\text{IMU}}}^2$ \cite{a13}. Additionally, the orientation data facilitates the alignment of LoS/NLoS azimuth angles across different UEs to mitigate the effects of varying orientations. Thus, it enables the consideration of azimuth angles for all $J$ UEs in the same coordinate system, as outlined in (\ref{e3}) and (\ref{e4}).
The Gaussian noise model of these measurements is selected for the sake of universality. Meanwhile, other measurement models can be easily adapted for the proposed algorithm.

Furthermore, we equip a stereo camera with a fixed location and orientation at the BS side. This camera is tasked with identifying targets within the BS's coverage area and conducting multi-object localization. This process involves utilizing depth information from the camera in conjunction with detections made by the YOLO algorithm \cite{YOLO}. For instance, at time $t$, the camera detects $N(t)$ targets within the RGB image, identified by their pixel coordinates $[u_{n'}(t), v_{n'}(t)]$, $n' = 1, \ldots, N(t)$. These detections are then converted into position estimates ${{{\hat{ \mathbf x}}_{{\text{VIS}},n'}}( t )=[ { \hat x}_{{\text{VISX}},n'}( t ), { \hat x}_{{\text{VISY}},n'} ( t )]}$, $n' = 1, \ldots,N( t )$, leveraging both the camera's depth image and orientation data \cite{MOL}.

A comprehensive overview of the proposed cooperative radio map construction, UE localization, and beam management algorithm is illustrated in Fig.~\ref{f2}. The framework consists of three interconnected sub-algorithms: multi-user SLAM, multi-modal enhanced localization, and sensing-aided beam management.
Initially, multi-user SLAM exploits radio and IMU measurements to perform local SLAM at the UE side, facilitating simultaneous multi-user localization and partial radio map construction. The BS subsequently integrates local radio maps and refines the global map using a Bayesian framework via a two-stage SLAM algorithm, further discussed in Sections~III and IV.
Next, vision measurements from the BS camera enhance the localization accuracy obtained via multi-user SLAM. The fusion of vision, radio, and IMU data supports multi-modal enhanced localization, as detailed in Section~V.
Finally, sensing-aided beam management utilizes the global radio map and UE localization results to generate priors for beam selection and tracking. These priors assist in identifying optimal beamforming vectors and constructing beam tracking codebooks, as elaborated in Section~VI.

\begin{figure}
\centering
\includegraphics[scale=0.45]{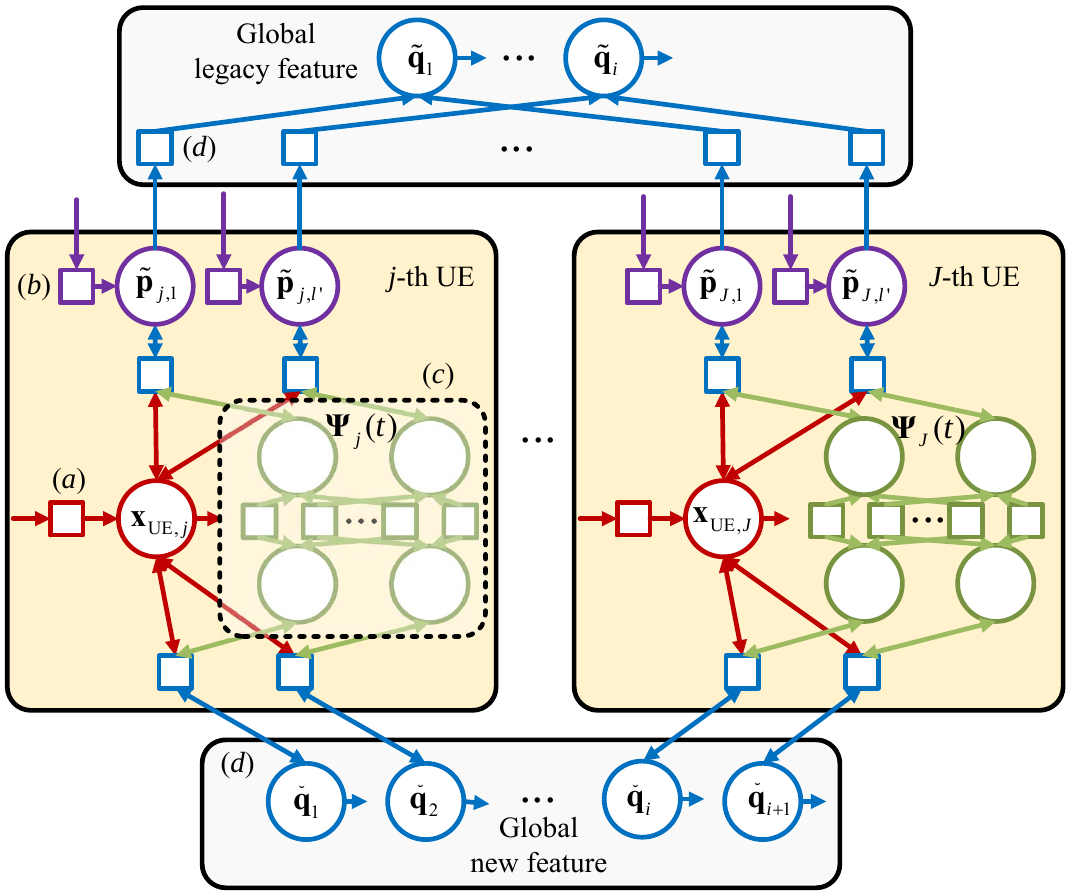}
\caption{ Multi-user SLAM factor graph.}
\label{f3}\vspace{-2mm}
\end{figure}

\section{Bayesian Framework for Multi-user SLAM}\label{III}
Multi-user SLAM focuses on gradually constructing the radio map composed of multiple VAs and locating UEs continuously utilizing radio measurements. 
This section sets the groundwork for the development of a practical SLAM algorithm by outlining a Bayesian estimation framework designed to tackle the multi-user SLAM problem. 
The essence of this framework lies in deducing various states from the collective measurement state ${{\mathbf{z}}_{1:J}}( {1:T} )$, gathered from all UEs over time. Specifically, the measurement state for the $j$-th UE at time $t$ is based on (\ref{e7}) and is represented as
\begin{equation}
{{\bf{z}}_j}(t) = {{\left[ {{{\hat \theta }_{j,1}}(t),{{\hat \phi }_{j,1}}(t), \ldots ,{{\hat \theta }_{j,{\hat L_j}(t)}}(t),{{\hat \phi }_{j,{\hat L_j}(t)}}(t)} \right]}^{\text{T}}}.
\end{equation}
The estimation is categorized into two main areas: local SLAM states and the global radio map state. For the local SLAM state, UEs autonomously determine their positions and map the immediate environment. Conversely, the global radio map state involves the BS or a designated central unit amalgamating the individual SLAM states from all UEs to create a unified global radio map.

The local SLAM states comprise three elements:
\begin{itemize}
\item {\bf UE Location:} The position of the UE, denoted as ${{\mathbf{x}}_{{\text{UE}},j}}( {t} )$.

\item {\bf Local Radio Map Set:} Denoted as ${{\mathbb {P}}_j}( t )$, it includes multiple local feature components ${{\bf{p}}_{j,l'}}$ represented by
\begin{equation}
\label{e10}
{{\bf{p}}_{j,l'}}= [ {{{\bf{ x}}_{{\text{VA}},j,l'}},{{ r}_{j,l'}}} ], ~l'=1,\ldots, {\left| {{\mathbb {P}}_j}(t) \right|},
\end{equation}
where the augmented vector ${{\bf{p}}_{j,l'}}$ combines the feature location ${{{\bf{ x}}}_{{\text{VA}},j,l’}}=[ {{{ x}_{{\text{VAX}},j,l’}},{{ x}_{{\text{VAY}},j,l’}}} ]$ and the confidence ${{ r}_{j,l’}}\in (0,1)$ indicating the probability of feature existence. Due to the dynamic nature of the environment, ${ {\mathbb {P}}_j}( t ) ={{\tilde { \mathbb {P}}}_j}( t ) \cup {{\breve {\mathbb {P}}}_j}( t ) $ is divided into legacy features ${{\tilde {\mathbb {P}}}_j}( t )$ and newly observed features ${{\breve {\mathbb {P}}}_j}( t ) = \{ \breve{\bf{p}}_{j,l’} \}$.

\item {\bf Data Association Vector:} ${{\mathbf{\Psi}} _{j}}( {t} )$, which links measurements to local features, as modeled in \cite{a2}.
\end{itemize}

For constructing the global radio map, we define the global radio map set ${{\mathbb {Q}}}( t )$ as an aggregation of multiple global feature components ${{\bf{q}}_{i}}$. Each global feature component can be expressed as an augmented vector, given by
\begin{equation}
\label{e11}
{{\bf{q}}_i} = {\left[ {{{\bf{x}}_{{\text{VA}},i}},{{\bf{v}}_i},{r_i},{t_i}} \right]}, ~i = 1,\ldots, | {\mathbb {Q}}(t) |.
\end{equation}
In this formulation, the feature location is given by ${{\bf{x}}_{{\text{VA}},i}}= [{{x}_{{\text{VAX}},i}},{{x}_{{\text{VAY}},i}}]$, and the location estimation covariance matrix is defined as ${{\bf{\Sigma}}_{{\text{VA}},i}^2} ={\text{diag}}( {\sigma _{{\text{VAX}},i}^2},{\sigma _{{\text{VAY}},i}^2})$. Moreover, ${{\bf{v}}_i} \in {\mathbb{R}^{1 \times J}}$ is a visibility vector in which each element ${v_i}[j]\in (0,1)$ indicates the likelihood of the $i$-th global feature contributing to a path in the $j$-th UE's channel $\mathbf{H}_{j}{(t)}$. 
The confidence level $r_i \in (0,1)$ represents the probability of the feature's existence, and $t_i$ denotes the timestamp marking the most recent update of the global feature.

\begin{figure*}[tb]{\color{black}\small
\begin{equation}\label{e000}
\begin{aligned}
&f\bigg( {\mathbb{Q}{\left( {0:T} \right)},{{\mathbf{x}}_{{\text{UE}},1:J}}{\left( {0:T} \right)},{\mathbb{P}_{1:J}}{\left( {0:T} \right)},{{\mathbf{\Psi}} _{1:J}}{\left( {1:T} \right)}\bigg|{{\mathbf{z}}_{1:J}}{\left( {1:T} \right)}} \bigg)\\
&\propto 
f\bigg( {\mathbb{Q}}{\left( {1:T} \right)},{{\mathbf{x}}_{{\text{UE}},1:J}}{\left( {1:T} \right)},{\mathbb{P}_{1:J}}{\left( {1:T} \right)},{{\mathbf{\Psi}} _{1:J}}{\left( {1:T} \right)} \bigg)
f\bigg({{\mathbf{z}}_{1:J}}{\left( {1:T} \right)}\bigg|{{\mathbf{x}}_{{\text{UE}},1:J}}{\left( {1:T} \right)},{\mathbb{P}_{1:J}}{\left( {1:T} \right)},{{\mathbf{\Psi}} _{1:J}}{\left( {1:T} \right)}  \bigg)
\\
& = \prod_{t = 1}^T {\left[ f\bigg( {\mathbb{Q}(t)\bigg{|}\mathbb{Q}{\left( {t - 1} \right)},{\mathbb{P}_{{1:J}}}(t)} \bigg) \prod_{j = 1}^J
f\bigg( {{{\mathbf{x}}_{{\text{UE}},j}}(t),{\mathbb{P}_j}(t),{\mathbf{\Psi} _j}(t)\bigg|{{\mathbf{x}}_{{\text{UE}},j}}(t-1),\mathbb{Q}{\left( {t - 1} \right)}},{{\mathbb{P}}_j}{\left( t-1 \right)} \bigg)
f\bigg( {{{\mathbf{z}}_j}(t)\bigg|{{\mathbf{x}}_{{\text{UE}},j}}(t),{\mathbb{P}_j}(t),{\mathbf{\Psi} _j}(t)} \bigg)
\right]}\\
& \propto 
\prod_{t=1}^{T}
\Bigg[
\underbrace{f\left( \mathbb{Q}(t) \bigg| \mathbb{Q}(t{-}1), \mathbb{P}_{1:J}(t) \right)}_{\text{(d) Global map update}} 
\cdot \prod_{j=1}^{J}
\Bigg(
\underbrace{f\left( \mathbf{x}_{\text{UE},j}(t) \bigg| \mathbf{x}_{\text{UE},j}(t{-}1) \right)}_{\text{(a) UE motion prediction}} 
\cdot
\underbrace{f\left( \tilde{\mathbb{P}}_j(t) \bigg| \mathbb{P}_j(t{-}1), \mathbb{Q}(t{-}1) \right)}_{\text{(b) Local map transition}} \\
& \hspace{5.5cm}
\cdot
\underbrace{f\left( \mathbf{z}_j(t) \bigg| \mathbf{x}_{\text{UE},j}(t), \mathbb{P}_j(t), \mathbf{\Psi}_j(t) \right)
f\left( \breve{\mathbb{P}}_j(t), \mathbf{\Psi}_j(t) \bigg| \tilde{\mathbb{P}}_j(t), \mathbf{x}_{\text{UE},j}(t) \right)}_{\text{(c) Measurement update and data association}}
\Bigg)
\Bigg] 
\end{aligned}
\end{equation}\vspace{-2mm}
\hrule
}
\end{figure*}

After specifying the states within the multi-user SLAM framework, we derive the joint posterior probability density function (PDF) for these states over the time period $t\in [0, T]$ for all $J$ UEs, as shown in (\ref{e000}). 
Initially, we assume the radio map states as ${\mathbb{Q}}( 0)=\emptyset$, ${\mathbb{P}_{1:J}}=\emptyset$. The PA location is known for VA calculation.
The initial UE state is estimated as ${\hat{\mathbf{x}}_{{\text{UE}},j}}( 0) \sim\mathcal{N}_2{({{\mathbf{x}}_{{\text{UE}},j}}( 0),{\bf{\Sigma}} _{{\text{ini}}})}$, where ${\bf{\Sigma}} _{{\text{ini}}} ={\text{diag}}( {\sigma _{{\text{ini}}}^2},{\sigma _{{\text{ini}}}^2})$.
Therefore, the joint posterior PDF can be obtained using Bayes' theorem, as indicated in the second line of (\ref{e000}). Next, by invoking the Markov state-transition principle, the joint posterior PDF is decomposed into PDFs for each UE $j$ at each time $t$, as shown in the third line of (\ref{e000}).
Finally, the joint posterior PDF is partitioned into four distinct segments \cite{a2} at any given time $t$:

\noindent \textbf{(a) State-transition PDF of UE Motion:}
This segment, $f( {{{\mathbf{x}}_{{\text{UE}},j}}( t )|{{\mathbf{x}}_{{\text{UE}},j}}( {t - 1} )} )$, encapsulates the state-transition PDF for the UE location ${{\mathbf{x}}_{{\text{UE}},j}}( t )$, predicated on its previous state ${{\mathbf{x}}_{{\text{UE}},j}}( t-1 )$. It characterizes the UE's process of predicting its preliminary location.

\noindent \textbf{(b) State-transition PDF of Local Radio Map:}
The next segment, $f( {{\tilde{\mathbb{P}}_j}( t )|{{\mathbb{P}}_j}( t-1 ),\mathbb{Q}( {t - 1} )} )$, details the state-transition PDF for the local radio map ${\tilde{\mathbb{P}}_j}( t )$, relying on its former state ${\tilde{\mathbb{P}}_j}( t-1 )$ and the prior global radio map $\mathbb{Q}( {t - 1} )$. This illustrates the UE's method of incorporating the global radio map to update its local map.

\noindent \textbf{(c) Likelihood Function and Measurement Update:}
This part incorporates the likelihood function for ${\mathbf{z}}_{j}(t)$ alongside the posterior PDF for ${{\breve {\mathbb {P}}}_j}( t )$ and ${{\mathbf{\Psi}} _{j}}( {t} )$. It delineates the UE's procedure of evaluating measurements against features, conducting data association, and refreshing feature states based on radio measurements.

\noindent \textbf{(d) Posterior PDF of Global Radio Map:}
Lastly, $f( {\mathbb{Q}( t )|\mathbb{Q}( {t - 1} ),{\mathbb{P}_{{1:J}}}( t )} )$ defines the posterior PDF for the global radio map ${\mathbb{Q}}( t )$, derived from its preceding state ${\mathbb{Q}}( {t - 1} )$ and the collective local radio maps ${\mathbb{P}}_{1:J}( t )$. This segment expounds on how the BS compiles the global radio map from the aggregated local maps.

This structured approach underpins the Bayesian estimation framework for multi-user SLAM, as visualized in the factor graph in Fig. \ref{f3}, with the components on the UE side shown in yellow and those on the BS side shown in gray. Four distinct segments, (a), (b), (c), and (d), are represented in red, purple, green, and blue, respectively \cite{b0}.
Initially, the radio maps $\mathbb{Q}( 0 ), \mathbb{P}_j( 0 )$ are assumed to be empty sets, in contrast to the more robust global radio map developed over time.

\section{Two-stage Multi-user SLAM Algorithm}\label{IV}
Building on the Bayesian framework introduced earlier, this section delineates a two-stage algorithm for multi-user SLAM utilizing a unified factor graph in Fig. \ref{f3}. 
The algorithm is divided into an initialization stage and a refinement stage.
Initially, a global radio map is constructed. UEs autonomously engage in local SLAM using the previous local radio map as a prior, and the BS formulates the initial global radio map based on the local SLAM outcomes from UEs. In the refinement phase, UEs gain access to the global radio map, facilitating further enhancements by both UEs and the BS. These stages are elaborated upon in the sections that follow.

\subsection{Initialization Stage}
This phase commences with UEs performing local SLAM in the absence of a global radio map. The results are then forwarded to the BS for the initial construction of the global radio map. Given the absence of $\mathbb{Q}( t )$ for UEs during this stage, component (b) in (\ref{e000}) is adjusted accordingly:
\begin{equation}
f{{\left( {\tilde{\mathbb{P}}_j}( t )\Big| {{\mathbb{P}}_j}( t-1 ),\mathbb{Q}( {t - 1} ) \right)}}=f{{\left( {\tilde{\mathbb{P}}_j}( t )\Big| {{\mathbb{P}}_j}( t-1 ) \right)}}.
\end{equation}
Hence, local SLAM, comprising components (a)--(c) in (\ref{e000}) but lacking a global map, is executed utilizing belief propagation through particle filters \cite{a2}. The local feature estimates ${\hat {\bf x}_{{\text{VA}},j,l'}}(t)=[{\hat { x}_{{\text{VAX}},j,l'}}(t),{\hat { x}_{{\text{VAY}},j,l'}}(t)]$ alongside the UE's location estimate ${\hat {\bf x}_{{\text{{RF\&IMU}}},j}}(t)=[{\hat { x}_{{\text{{RF\&IMU}X}},j}}(t),{\hat { x}_{{\text{{RF\&IMU}Y}},j}}(t)]$ from local SLAM are based on IMU predictions and angle estimates in (\ref{e7}). The IMU predictions realize the state transition of UE location in component (a), and the angle estimates realize the measurement update in component (c). The feature estimates and UE locations are assumed to follow two-dimensional Gaussian distributions: 
\begin{subequations}\label{e14}
\begin{align}
{\hat {\bf x}_{{\text{VA}},j,l'}} 
& \sim{\mathcal N}_2
{\left(
{\hat {\bf x}_{{\text{VA}},j,l'}} ,
{\bf{\Sigma}} _{{\text{VA}},j,l'}
\right)},  \\
{\hat {\bf x}_{{\text{{RF\&IMU}}},j}}  & \sim{\mathcal N}_2
{\left(
{\hat {\bf x}_{{\text{{RF\&IMU}}},j}},
{\bf{\Sigma}} _{{\text{RF\&IMU}},j}
\right)}, \label{e14cd}
\end{align}
\end{subequations}
with covariance matrixes ${\bf{\Sigma}} _{{\text{VA}},j,l'}={\text{diag}}(\sigma _{{\text{VAX}},j,l'}^2 ,\sigma _{{\text{VAY}},j,l'}^2)$, ${\bf{\Sigma}} _{{\text{{RF\&IMU}}},j}={\text{diag}}(\sigma _{{\text{{RF\&IMU}X}},j}^2,\sigma _{{\text{{RF\&IMU}Y}},j}^2)$. 
Time $t$ is omitted for brevity. 
Subsequently, the local feature data from all $J$ UEs are collated and sent to the BS.

At the BS side, initializing the global radio map equates to estimating the BS state, achieved through designing part $(d)$. The process is delineated with inputs $\mathbb{Q}(t - 1)$ and ${\mathbb{P}_{{j}}}( t )$, and yields the output $ \mathbb{Q}( t )$. 
Initially, we define ${\mathbb{P}_{\text{pool}}}( {t } )$ to encompass the pruned local VAs, as specified by
\begin{equation}
\label{e15}
{{\mathbb{P}}_{{\text{pool}}}}(t) = \bigcup\limits_{j = 1}^J \left \{ {{\bf p}_{j,l'}} \Big |{{\bf p}_{j,l'}} \in {{\mathbb{P}}_j}(t),~{r_{j,l'}} \geq {r_{{\text{cut}}}} \right \},
\end{equation}
where $r_{{\text{cut}}}$ serves as the pruning threshold. The evolution of the global radio map ${{\mathbb {Q}}}(t)$ is based on ${{\mathbb {Q}}}(t-1)$ and ${\mathbb{P}_{\text{pool}}}( {t } )$.

Next, we proceed by evaluating the features in ${\mathbb{P}_{\text{pool}}}( {t } )$ for data association with ${{\mathbb {Q}}}(t-1)$. For each local feature ${{\bf p}_{j,l'}} = {\left[ {{\hat {\bf x}_{{\text{VA}},j,l'}},{r_{j,l'}}} \right]}$, we identify the nearest global feature in ${{\mathbb {Q}}}(t-1)$ to ${{\bf p}_{j,l'}}$, denoted as ${{\bf q}_{{\text{cand}}}}$. The proximity, calculated as ${{d}} = {\| {{\hat {\bf x}_{{\text{VA}},j,l'}} - {\hat {\bf x}_{{\text{VA}},{\text{cand}}}}} \|_2}$, is assessed against a set threshold ${d_{{\text{th}}}}$. Should ${{d}}\leq {d_{{\text{th}}}}$, we acknowledge a successful match and accordingly combine ${{\bf p}_{j,l'}}$ in ${\mathbb{P}_{\text{pool}}}( {t } )$ with ${{\bf q}_{{\text{cand}}}}$ in ${{\mathbb {Q}}}(t-1)$ by applying
\begin{subequations}
\label{e16}
\begin{align}
{\hat x_{{\text{VAX/Y}},{\text{cand}}}} &= \frac{{\sigma _{{\text{VAX/Y}},j,l'}^2 {\hat x_{{\text{VAX/Y}},{\text{cand}}}} + \sigma _{{\text{VAX/Y}},{\text{cand}}}^2 {\hat x_{{\text{VAX/Y}},j,l'}}}}{{\sigma _{{\text{VAX/Y}},{\text{cand}}}^2 + \sigma _{{\text{VAX/Y}},j,l'}^2}}, \label{e16ab} \\
\sigma _{{\text{VAX/Y}},{\text{cand}}}^2 &= {{\left( {\frac{1}{{\sigma _{{\text{VAX/Y}},{\text{cand}}}^2}} + \frac{1}{{\sigma _{{\text{VAX/Y}},j,l'}^2}}} \right)}^{ - 1}} ,\label{e16cd} \\
{{\mathbf{v}}_{{\text{cand}}}}{\left[ j \right]} &= {r_{j,l'}} , \label{e16e}\\
{r_{{\text{cand}}}} &= \max \left\{ {{{\mathbf{v}}_{{\text{cand}}}}} \right\}, \label{e16f} \\
{t_{{\text{cand}}}} &= t \label{e16g}.
\end{align}
\end{subequations}
In these equations, ${\hat {\bf x}_{{\text{VA}},{\text{cand}}}}$ is updated by fusing the measurements ${\hat {\bf x}_{{\text{VA}},j,l'}}$ and the prior ${\hat {\bf x}_{{\text{VA}},{\text{cand}}}}$ according to (\ref{e16ab}), and the covariance is updated using (\ref{e16cd}). 
Equation (\ref{e16e}) refreshes the $j$-th index of the visibility vector ${\mathbf{v}}_{\text{cand}}$ based on the local feature's confidence ${r_{j,l'}}$, while (\ref{e16f}) and (\ref{e16g})  refresh the global feature's confidence and timestamp, respectively. (Note that the subscript $(\cdot)_{\text{VAX/Y}}$ is used for brevity; the values for $(\cdot)_{\text{VAY}}$ are inferred similarly by replacing $(\cdot)_{\text{VAX}}$.) 
If ${{d}} > {d_{{\text{th}}}}$, indicating an unsuccessful association, the local feature ${{\bf p}_{j,l'}}$ is classified as a new global feature ${{\bf q}_{i}}$, inheriting the same location and variance properties. The corresponding attributes, including the visibility vector, confidence, and timestamp, are then assigned analogously to (\ref{e16e}) through (\ref{e16g}). This new global feature ${{\bf q}_{i}}$ is subsequently incorporated into ${\mathbb{Q}}(t)$. 

Following the examination of ${\mathbb{P}_{\text{pool}}}( {t } )$, the process entails updating ${{\mathbb {Q}}}(t-1)$ with features that have successfully associated and forming ${{\mathbb {Q}}}(t)$ with features lacking successful associations. This results in the amalgamation of both to establish the refined global radio map, denoted as ${{\mathbb {Q}}}(t)={{\mathbb {Q}}}(t)\cup {{\mathbb {Q}}}(t-1)$. Note that features from ${{\mathbb {Q}}}(t-1)$ may persist even in the absence of successful associations, thereby necessitating a pruning mechanism that takes into account both the feature's timestamp ${t_{\text{cand}}}$ and its confidence level ${r_{\text{cand}}}$. The pruning criterion is articulated as follows
\begin{equation}
\label{e17}
{{\mathbb {Q}}}(t) = \left\{ {{\bf q}_{\text{cand}}} \Big|{{\bf q}_{\text{cand}}} \in {{\mathbb {Q}}}(t),~{2^{{t_{{\text{cand}}}} - t}} \cdot {r_{\text{cand}}} > {r_{{\text{cut}}}} \right\},
\end{equation}
where ${r_{{\text{cut}}}}$ is the threshold defined in (\ref{e15}). Post-pruning, the global radio map ${{\mathbb {Q}}}(t)$ is finalized. This stage is encapsulated in Algorithm \ref{A1}, summarizing the initial stage of the multi-user SLAM process.

This procedure marks the completion of the global radio map initialization, culminating the design for part (d). By integrating the approaches delineated for parts (a)–(c) on the UE side, the initialization phase of the multi-user SLAM is comprehensively achieved.
\begin{algorithm}[t]
\caption{Global Radio Map Construction in Initial Stage}\label{A1}
\KwIn{Global radio map ${\mathbb {Q}} {\left({t-1}\right)}$; Uploaded local feature set ${\mathbb {P}}_{\text{pool}} {\left({t}\right)}$;}
Set ${\mathbb {Q}} {\left({t}\right)}=\emptyset$\;
\For{\rm ${{\bf p}_{j,l'}} \in {\mathbb {P}}_{\text{pool}} {\left({t}\right)}$ }
{
Search for ${{\bf q}_{{\text{cand}}}} = \mathop {{\text{argmin}}}\limits_{{{\bf q}_{{\text{cand}}}} \in {\mathbb Q}{\left( {t - 1} \right)}} {{ {{\| {{\hat{\bf{x}}_{{\text{VA}},j,l'}} - {\hat{\bf{x}}_{{\text{VA}},{\text{cand}}}}} \|}} }_2}$\;
Calculate $d = { {\| {\hat{\bf{x}}_{{\text{VA}},j,l'}} - {\hat{\bf{x}}_{{\text{VA}},{\text{cand}}}} \| }_2}$\;

\eIf{\rm${d}\leq {d_{{\text{th}}}}$}
{Update ${\bf q}_{{\text{cand}}}$ in ${\mathbb {Q}} {\left({t-1}\right)}$ based on (\ref{e16})\;

}
{Transfer ${{\bf p}_{j,l'}}$ into ${\bf q}_{i}$\;
Add ${\bf q}_{i}$ to ${\mathbb {Q}} {\left({t}\right)}$\;
}
}
Set ${\mathbb Q}(t) = {\mathbb Q}{\left( {t - 1} \right)} \cup {\mathbb Q}(t)$\;

Perform feature pruning for ${\mathbb Q}(t)$ based on (\ref{e17})\;
\KwOut{Global radio map ${\mathbb {Q}} {\left({t}\right)}$;}
\end{algorithm}

\subsection{Refinement Stage}
During the initial phase, UEs operate without access to the global radio map, hindering their ability to refine local SLAM activities. Moreover, the method of associating global and local radio maps through distance metrics may introduce discrepancies. These challenges stem from the structural designs of components (b) and (d) in (\ref{e000}). To overcome these hurdles, the refinement stage is introduced, tailored to recalibrate parts (b) and (d) upon the availability of the global radio map.

Firstly, part (b) undergoes modifications to facilitate the downloading of the global radio map by UEs for enhancing the local SLAM process. Specifically, UEs acquire ${{\mathbb {Q}}}(t-1)$ and construct ${\tilde{\mathbb{P}}_j}( t )$ utilizing the visibility vector ${\bf{v}}_i$ of the $i$-th global feature. Each global feature ${\bf{q}}_i$ within ${{\mathbb {Q}}}(t-1)$ is then designated as a local legacy feature ${\tilde{ \bf{p}}}_{j,i}$ within ${\tilde{\mathbb{P}}_j}( t )$ for the $j$-th UE, depicted as\footnote{Notably, we refer to ${\tilde{ \bf{p}}}_{j,l'}$ as the $l'$-th local legacy feature of the $j$-th UE. The only exception is in (\ref{e18}), where the $i$-th global feature ${\bf{q}}_i$ within ${{\mathbb {Q}}}(t-1)$ is downloaded as the $i$-th local legacy feature ${\tilde{ \bf{p}}}_{j,i}$ of the $j$-th UE. }
\begin{equation}
\label{e18}
{\tilde{\mathbf{p}}_{j,i}} = \left\{ \begin{gathered}
{ {\left[ {{\tilde{\mathbf{x}}_{{\text{VA,}}j,i}},{\tilde{r}_{j,i}}} \right]}={\left[ {{\hat{\mathbf{x}}_{{\text{VA,}}i}},{r_i}} \right]},{\text{\;\;\;\;if\;}}{{\mathbf{v}}_i}{\left[ j \right]} > {r_{{\text{cut}}}} },\hfill\\
{ {\left[ {{\tilde{\mathbf{x}}_{{\text{VA,}}j,i}},{\tilde{r}_{j,i}}} \right]}={\left[ {{\hat{\mathbf{x}}_{{\text{VA,}}i}},{r_{{\text{cut}}}}} \right]},{\text{\;\;otherwise}}}.\hfill \\
\end{gathered} \right.
\end{equation}
By iterating over all global features in ${{\mathbb {Q}}}(t-1)$, UEs compile ${\tilde{\mathbb{P}}_j}( t )$. This establishes the methodology for $f( {\tilde{\mathbb{P}}_j}( t )|f( {{\mathbb{P}}_j}( t -1),\mathbb{Q}( {t - 1} ) )=f( {\tilde{\mathbb{P}}_j}( t )|\mathbb{Q}( {t - 1} ) )$.
Consequently, UEs engage in local SLAM leveraging the updated ${\tilde{\mathbb{P}}_j}( t )$. Owing to the process outlined in (\ref{e18}) preserving features regardless of confidence levels, the sets of local legacy features, denoted as ${{{\tilde {\mathbb P}}_j}(t)}$ for $j=1,\ldots, J$, maintain consistent VA numbers and indices.

Subsequently, part (d) is refined to further enhance the global radio map. The process begins by categorizing the uploaded local radio maps into two distinct sets:
\begin{subequations}  
\begin{align}
\label{e19a}
\tilde{\mathbb{P}}_{\text{pool}}(t) &= \bigcup_{j=1}^J \left\{{\tilde{\mathbf{p}}_{j,l'}}={\left[ {\tilde{\mathbf{x}}_{{\text{VA,}}j,l'}},\tilde{r}_{j,l'} \cdot 1_{{\left(\tilde{r}_{j,l'} > r_{\text{cut}}\right)}} \right]} \right\}, 
\\
\breve{\mathbb{P}}_{\text{pool}}(t) &= \bigcup_{j=1}^J \left\{ \breve{\mathbf{p}}_{j,l'} \,\middle|\, \breve{\mathbf{p}}_{j,l'} \in \breve{\mathbb{P}}_j(t), ~{\breve r}_{j,l'} \geq {\breve r}_{\text{cut}} \right\},\label{e19b}
\end{align}
\end{subequations}

\noindent where set ${\tilde {\mathbb{P}}_{{\text{pool}}}}( t )$ consists of legacy local features, and ${\breve {\mathbb{P}}_{{\text{pool}}}}( t )$ consists of new local features.
The construction of ${\tilde {\mathbb{P}}_{{\text{pool}}}}( t )$ adjusts feature confidence $\tilde{r}_{j,l'}$ without pruning any features. Hence, the local feature index remains consistent across UEs.
Therefore, legacy global features can be updated similarly to the successful data association case in the initial stage. For instance, the $i$-th legacy global feature $\tilde{\bf{q}}_i$ can be updated using the local features $\tilde{r}_{j,i},\,j=1,\ldots,J$.
The construction of ${\breve {\mathbb{P}}_{{\text{pool}}}}( t )$ in (\ref{e19b}) is performed similarly to (\ref{e15}). Subsequently, local new features in ${\breve {\mathbb{P}}_{{\text{pool}}}}( t )$ are transformed into global new features.
Finally, the updated legacy global features and new global features can form the global radio map ${\mathbb{Q}}(t)$. This process completes the adjusted design for part (d). With these adjustments, the refinement stage can be implemented, which is summarized in Algorithm \ref{A2}. This stage requires the global radio map to contain reliable global features for UEs to download. 
Integrating different sensors at both the BS and UE sides with the proposed multi-user SLAM algorithm holds significant promise for advancing UE localization capabilities. In this work, we consider the multi-modal enhanced UE localization based on measurements from a stereo camera at the BS and IMUs at the UEs, which is introduced in the next section. 

\begin{algorithm}[t]
\caption{Global Radio Map Construction in Refinement Stage}\label{A2}
\KwIn{Global Map ${\mathbb {Q}} {\left({t-1}\right)}$\;
\:\:\:\:\:\:\:\:\:\:\:\:\:\:Uploaded legacy local feature set ${\tilde {\mathbb P} _{{\text{pool}}}}(t)$\;
\:\:\:\:\:\:\:\:\:\:\:\:\:\:Uploaded new local feature set ${\breve {\mathbb P} _{{\text{pool}}}}(t)$\;}

Set ${\tilde{\mathbb Q}} {\left({t}\right)}={{\mathbb Q}} {\left({t-1}\right)}$\;
\For {${\tilde {\bf q}_i} \in {\tilde {\mathbb Q}}(t)$}
{

Set $j=1$\;
\While {$j \leq J$}
{
Update ${{{\tilde {\bf v}}_i}}{\left[ j \right]} = {{\tilde r}_{j,i}}$\;
\If{ \rm${{{\tilde {\bf v}}_i}}{\left[ j \right]} > 0$ }
{
Update ${\tilde {\bf q}_i}$ utilizing ${\tilde {\bf p}_{j,i}}$ similar to (16)\;
}
$j=j+1$;
}
}

Set ${\breve {\mathbb Q}}(t) = \emptyset $\;
\For{\rm${\breve {\bf p} _{j,l'}} \in {\breve {\mathbb P} _{{\text{pool}}}}(t)$}
{
Search ${\breve {\bf q} _{{\text{cand}}}} = \mathop{{\text{argmin}}}\limits_{{{\breve {\bf q} }_{{\text{cand}}}} \in {{\breve {\mathbb {Q}} }}(t)} {{ {{\| {{{\breve {\bf x} }_{{\text{VA}},j,l'}} - {{\breve {\bf x} }_{{\text{VA,cand}}}}} }} \|}_2}$\;
Calculate ${d} = {{ {{\| {{{\breve {\bf x} }_{{\text{VA}},j,l'}} - {{\breve {\bf x} }_{{\text{VA,cand}}}}} \|}} }_2}$\;
\eIf{\rm${d} \le {d_{{\text{th}}}}$}
{
Update ${\breve {\bf q} _{{\text{cand}}}}$ in ${\breve {\mathbb {Q}}}(t)$ similar to (16)\;
}
{Transfer ${\breve {\bf p}_{j,l'}}$ into global feature $\breve{\bf q}_{i}$\;
Add $\breve{\bf q}_{i}$ to ${\breve{\mathbb {Q}}} {\left({t}\right)}$\;
}
}
Set ${\mathbb {Q}} {\left({t}\right)}={\breve {\mathbb {Q}} }(t) \cup {\tilde {\mathbb {Q}} }(t) $\;
Perform feature pruning for ${\mathbb Q}(t)$ based on (\ref{e17})\;
\KwOut{Global radio map ${\mathbb {Q}} {\left({t}\right)}$;}
\end{algorithm}

{\textit{
Remark 1:
The initialization and refinement stages of the algorithm exhibit distinct characteristics. Specifically, the initialization stage is crafted to establish the global radio map independently, without any prerequisite of its prior existence. In contrast, the refinement stage actively utilizes the global radio map, thus necessitating its prior establishment and reliability. In the context of our design, the initialization phase is set to occur within the time frame {\rm$0 < t \leq T_{\text{config}}$}, which then transitions to the refinement phase for {\rm$t > T_{\text{config}}$}.
}}

\section{Multi-modal Enhanced localization}\label{V}
The two-stage multi-user SLAM algorithm primarily focuses on the collaborative construction of the global radio map $\mathbb{Q}(t)$. Although the local SLAM processes, based on parts (a)--(c), facilitate UE localization, it is noteworthy that the refinement stage, which leverages the now-available $\mathbb{Q}(t)$, yields only marginal improvements in localization accuracy compared to scenarios involving a single UE. This section addresses the integration of the multi-user SLAM localization results, ${{{\hat{ \mathbf x}}}_{{\text{{RF\&IMU}}},j}}( t )$ from (\ref{e14cd}), with vision-based localization results, ${{{\hat{\mathbf x}}}_{{\text{VIS}},n'}}( t ) $, to achieve multi-modal enhanced localization using radio, IMU, and vision measurements. 

Initially, we formulate a probability model for the outcomes of camera-based multi-object localization. Subsequently, we define the process of data association between $J$ UEs and $N(t)$ objects identified by the camera system. The final part of this section is dedicated to discussing how the localization results of the associated pairs are integrated.

Given that the detection and localization process for each object is independent, we postulate that the location estimates of objects adhere to Gaussian distributions. Considering the $n'$-th object ${{\hat{\mathbf x}}_{{\text{VIS}},n'}}(t)$, which is successfully associated with the $j$-th UE, the following distributions are proposed 
\begin{equation}\label{e20}
{{\hat{\mathbf x}}_{{\text{VIS}},n'}} \sim
\mathcal{N}_2{\left(
{{\hat{\mathbf x}}_{{\text{VIS}},n'}},
{\bf{\Sigma}} _{{\text{VIS}},n'}
\right)},
\end{equation}
where ${\bf{\Sigma}} _{{\text{VIS}},n'}={\text{diag}}(\sigma _{{\text{VISX}}}^2 ( {{u_{n'}},{v_{n'}}} ) ,\sigma _{{\text{VISY}}}^2 ( {{u_{n'}},{v_{n'}}} ))$ represents the covariance matrix describing the object localization error.
The temporal variable $t$ is presumed to be implicit for conciseness. 
Given the camera's fixed location and orientation, the error associated with object localization is significantly influenced by the pixel coordinates $[{{u_{n'}},{v_{n'}}}]$ obtained through YOLO detection. Hence, the localization variance functions are modeled as $\sigma _{{\text{VISX}}}^2( {{u},{v}} )$ and $\sigma _{{\text{VISY}}}^2( {{u},{v}} )$, with ${{u=u_{n'}},{v=v_{n'}}}$, to accurately represent this relationship.

Subsequent to the offline training, we construct a lookup table that maps pixel coordinates to localization errors, effectively fitting the variance functions. This involves segmenting the entire image into blocks by dividing at each $u_{\text{cell}}$ horizontally and $v_{\text{cell}}$ vertically within the RGB image. The variance functions, in consideration of these blocks, are delineated as
\begin{equation}\label{e21}
\sigma _{{\text{VISX/Y}}}^2{\left( {u,v} \right)} = e_{{\text{VISX/Y}}}^2{\left( \left\lceil \frac{u} {u_{\text{cell}}} \right\rceil ,\left\lceil \frac{v}{v_{\text{cell}}}\right\rceil \right)},
\end{equation}
where $e_{{\text{VISX/Y}}}( \lceil \frac{u} {u_{\text{cell}}} \rceil,\lceil \frac{v}{v_{\text{cell}}}\rceil )$ represents the average estimation errors in coordinates for the block indexed by $\lceil \frac{u} {u_{\text{cell}}} \rceil $ and $\lceil \frac{v}{v_{\text{cell}}}\rceil $. Through this offline training process, the variance for the $n'$-th object is determined using its pixel coordinates $[{u_{n'}},{v_{n'}}]$ based on (\ref{e21}). An example implementation of this offline training is provided in Section VII-B.

In the subsequent steps, we finalize the data association and localization integration processes using the outcomes of offline training. Given the sparse distribution of UEs and objects in the physical space, data association is predicated on the proximity between the SLAM-based localization results ${{{\hat{ \mathbf x}}}_{{\text{{RF\&IMU}}},j}}( t )$ and the vision-based localization estimates ${{{\hat{\mathbf x}}}_{{\text{VIS}},n'}}( t )$. Iterating through all UEs to complete data association is essential, as the proposed localization algorithm focuses on UEs. For instance, for the $j$-th UE, we identify the closest object among $N(t)$ objects, denoted by the index ${{{I}}_{j}}$, as follows:
\begin{equation}
{{{I}}_{j}} = \mathop {{\text{argmin}}}\limits_{{{{I}}_{j}} \in \left\{ {1, \ldots ,N( t )} \right\}} { {\left\| {{{{\hat{\mathbf x}}}_{{\text{VIS}},{{{I}}_{j}}}} - {{{\hat{ \mathbf x}}}_{{\text{{RF\&IMU}}},j}}} \right\|}_2}.
\end{equation}

We then calculate the distance $D_{\text{VIS-{RF\&IMU}}}$ between the SLAM localization result ${{{\hat{ \mathbf x}}}_{{\text{{RF\&IMU}}},j}}( t )$ and the vision-based localization result ${{\hat{\bf x}}}_{{\text{VIS}},{{{I}}_{j}}}$, and the distance $D_{\text{ VIS-IMU}}$ between the vision-based localization result ${{\hat{\bf x}}}_{{\text{VIS}},{{{I}}_{j}}}$ and the IMU prediction ${{{\hat{\bf x}}}_{{\text{IMU}},j}}$.
If $\max(D_{\text{ VIS-{RF\&IMU}}},D_{\text{ VIS-IMU}})$ is less than the threshold $D_{\text{T}h}$, it signifies successful data association between the $j$-th UE and the ${I}_{j}$-th object. According to (\ref{e14cd}) and (\ref{e20}), the $j$-th UE and ${I}_{j}$-th object localization results can be regarded as Gaussian distributions, with radio, IMU, and vision measurements serving as the posterior. Since these measurements are independent, the posterior PDF of multi-modal enhanced UE localization ${{\hat{\bf x}}_{{\text{UE}},j}}( t )$ based on all three measurements can be viewed as the product of ${\hat {\bf x}_{{\text{{RF\&IMU}}},j}}   \sim{\mathcal N}_2{({\hat {\bf x}_{{\text{{RF\&IMU}}},j}},{\bf{\Sigma}}_{{\text{RF\&IMU}},j})}$ and ${{\hat{\mathbf x}}_{{\text{VIS}},{I}_{j}}} \sim \mathcal{N}_2{({{\hat{\mathbf x}}_{{\text{VIS}},{I}_{j}}},{\bf{\Sigma}} _{{\text{VIS}},{I}_{j}})}$. Thus, multi-modal enhanced UE localization ${{\hat{\bf x}}_{{\text{UE}},j}}( t )$ can be delineated as 
\begin{equation}\label{e23}
\begin{aligned}
{\hat x_{{\text{UEX/Y}},j}}(t) &= \frac{{\sigma _{{\text{{RF\&IMU}X/Y}},j}^2}}{{\sigma _{{\text{{RF\&IMU}X/Y}},j}^2 + \sigma _{{\text{VISX/Y}}}^2{\left( {{u_{{{{I}}_{j}}}},{v_{{{{I}}_{j}}}}} \right)}}}{\hat x_{{\text{VISX/Y}},{{{I}}_{j}}}} \\
&+\frac{{\sigma _{{\text{VISX/Y}}}^2{\left( {{u_{{{{I}}_{j}}}},{v_{{{{I}}_{j}}}}} \right)}}}{{\sigma _{{\text{{RF\&IMU}X/Y}},j}^2 + \sigma _{{\text{VISX/Y}}}^2{\left( {{u_{{{{I}}_{j}}}},{v_{{{{I}}_{j}}}}} \right)}}}{\hat x_{{\text{{RF\&IMU}X/Y}},j}}.
\end{aligned}
\end{equation}

If the distance exceeds $D_{\text{th}}$, data association for the $j$-th UE is considered unsuccessful, and ${{\hat{\bf x}}_{{\text{UE}},j}}( t )$ remains as ${{\hat{\bf x}}_{{\text{{RF\&IMU}}},j}}$ without integration. This methodology is encapsulated in Algorithm \ref{A3}. The vision-based localization results and SLAM localization results are integrated at the end of each time step. This approach is adopted because fusing radio and vision measurements at an earlier stage could encounter significant interference from multiple targets in the vision data, especially in multi-user scenarios. The proposed algorithm focuses on UE localization. Therefore, the emphasis is on associating UE localization results obtained from radio and IMU measurements with multi-target localization results derived from the stereo camera, rather than implementing joint data association between radio and vision localization. The integration of different sensor measurements can complement each other, providing an opportunity for multi-modal SLAM in multi-user scenarios, thereby enhancing localization capabilities. 



\begin{algorithm}[t]
\caption{Multi-modal Enhanced Localization}\label{A3}
\KwIn{ IMU prediction ${{\hat{\bf x}}_{{\text{IMU}},j}}$; SLAM localization results ${{\hat{\bf x}}_{{\text{{RF\&IMU}}},j}}$; Camera-based multi-object localization results ${{\hat{\bf x}}_{{\text{VIS}},n'}}$\;}

Set $j=1$\;
\While{$j\leq J$}
{
Calculate ${{{I}}_{j}} = \mathop {{\text{argmin}}}\limits_{{{{I}}_{j}} \in \left\{ {1, \ldots ,N(t)} \right\}} { {\left\| {{{{\hat{\bf x}}}_{{\text{VIS}},{{{I}}_{j}}}} - {{{\hat{\bf x}}}_{{\text{{RF\&IMU}}},j}}} \right\|}_2}$\;
Calculate ${{{D}}_{{\text{VIS}} - {\text{{RF\&IMU}}}}} = { {\left\| {{{{\hat{\bf x}}}_{{\text{VIS}},{{{I}}_{j}}}} - {{{\hat{\bf x}}}_{{\text{{RF\&IMU}}},j}}} \right\|} _2}$\;
Calculate ${{{D}}_{{\text{VIS}} - {\text{IMU}}}} = {\left\| {{{{\hat{\bf x}}}_{{\text{VIS}},{{{I}}_{j}}}} - {{{\hat{\bf x}}}_{{\text{IMU}},j}}} \right\|}_2 $\;
\eIf{\rm$\max\left\{D_{\text{ VIS-{RF\&IMU}}},D_{\text{ VIS-IMU}}\right\} < {{{D}}_{{\text{th}}}}$}
{
Calculate ${{\hat{\bf x}}_{{\text{UE}},j}}(t)$ based on (\ref{e23})\;
}
{
${{\hat{\bf x}}_{{\text{UE}},j}}(t) = {{\hat{\bf x}}_{{\text{{RF\&IMU}}},j}}$\;
}
Set $j=j+1$\;
}
\KwOut{Enhanced UE location estimation ${{\hat{\bf x}}_{{\text{UE}},j}}(t)$\;}
\end{algorithm}

\section{Sensing-aided Beam Management}\label{VI} 
The preceding sections have detailed the construction of the global radio map $\mathbb{Q}(t)$ and the formulation of multi-modal enhanced localization $\hat{\mathbf{x}}_{\text{UE},j}$. Building on these foundations, this section leverages the derived sensing results to enhance the efficiency of beam management. The process unfolds in two main stages: the prior information generation step and the beam selection step. In the prior information generation stage, the radio map and UE localization results are used to generate a set of available beam directions for each UE. Subsequently, the beam selection step identifies the appropriate beams and enables IUI-reduced beam management. 

Initially, we embark on the process of deriving prior information for all conceivable paths associated with the $J$ UEs. This is meticulously allocated to ensure each UE benefits from distinct prior data relevant to its unique path. Illustratively, we consider the $j$-th UE to demonstrate how this prior information is generated. In the global radio map ${\mathbb Q}(t)$, global features ${\bf q}_i$, where the visibility vector ${\bf v}_{i}[j]>r_{\text{cut}}$ in ${\mathbb Q}(t)$ are deemed to provide viable paths for the $j$-th UE. Assuming $\hat{L}_{j}(t)$ features meet these criteria for the $j$-th UE, we construct the prior information set ${{\mathbb U}_j}(t)$. This set comprises $\hat{L}_{j}(t)$ prior information vectors ${{\bf u}_{j,l}}(t)$, articulated as
\begin{equation}\label{e24}
{{\bf u}_{j,l}} = {\left[ \hat {\theta} _{j,l}’,\hat {\phi} _{j,l}’,{{{{S}}}_{j,l}},{{\hat {g}}_{j,l}’} \right]}, ~l = 1, \ldots ,{\hat {L}_j},
\end{equation}
where $t$ is assumed implicit for brevity. Each vector ${{\bf u}_{j,l}}$ encapsulates the path's angular coordinates $\hat {\theta} _{j,l}’$ and $\hat {\phi} _{j,l}’$, an angular search range ${{{{S}}}_{j,l}}$, and the path's gain ${{\hat {g}}_{j,l}’}$. The generation process is as follows:

\noindent {\bf Path Angles $\hat {\theta} _{j,l}'$ and $\hat {\phi} _{j,l}'$:} These are derived using (\ref{e3}) and (\ref{e4}), leveraging the known PA ${\bf x}_{\text{PA}}$, the global feature's location from ${\mathbb Q}(t)$, and the UE's estimated position ${{\hat{\bf x}}_{{\text{UE}},j}}$.

\noindent {\bf Azimuth Searching Range ${{{{S}}}_{j,l}}$:} Generated based on \cite{a13}, this parameter delineates the minimal angular range around $\hat {\theta} _{j,l}’$ and $\hat {\phi} _{j,l}’$ that is likely to encompass the actual path azimuth.

\noindent {\bf Path Gain ${{\hat {g}}_{j,l}’}$:} Calculated using (\ref{e3b}) and (\ref{e4c}), it factors in the path's angular coordinates, the global feature's location from ${\mathbb Q}(t)$, the UE's estimated location ${{\hat{\bf x}}_{{\text{UE}},j}}$.

For ease of reference, it is stipulated that the sequence of prior information vectors ${{\bf {u}}_{j,l}}$ is ordered such that ${{\mathbb U}_j}(t)$ satisfy that ${{\hat g}_{j,1}’} > {{\hat g}_{j,2}’} > \ldots > {{\hat g}_{j,{{\hat L}_j}}’}$.
Furthermore, channel estimation ${{\hat{\bf{H}}}_j}(t) $ is realized as
\begin{equation}
{{\hat{\bf{H}}}_j}(t) = \mathop \sum \limits_{l = 1}^{{\hat L_j}(t)} {\hat g_{j,l}’}{{\bf{a}}_{{\text{UE}}}} {{\left( {{\hat\theta’ _{j,l}}} \right)}} {\bf{a}}_{{\text{BS}}}^{\text{H}} {{\left( {{\hat\phi’ _{j,l}}} \right)}}.
\end{equation}

\begin{algorithm}[t]
\caption{Sensing-aided Beam Management}\label{A4}
\KwIn{UE location estimation ${{\hat{\bf x}}_{{\text{UE}},j}}(t)$; Global radio map ${{\mathbb Q}}(t)$}
Generate ${{\mathbb U}_j},j = 1, \ldots ,J$ based on (\ref{e24})\;
Set ${{\bf u}_{j,{\text{opt}}}}={{\bf u}_{j,1}},j = 1, \ldots ,J$\;
Calculate $\hat {{\text{SE}}}$ based on (\ref{e26}) and (\ref{e27})\;
Set $j=1$\;
\While{$j\leq J$}
{
Calculate $\theta_{{\text{IUI}},j}$ based on (\ref{e28})\;

\If{\rm$\theta_{{\text{IUI}},j}<\theta _{\text{th}}$}
{
Switch ${{\bf u}_{j,{\text{opt}}}}$ to the next path in ${{\mathbb U}_j}(t)$\;
Calculate $\hat {{\text{SE}}}'$ based on (\ref{e26}) and (\ref{e27})\;

\eIf{ \rm$\hat {{\text{SE}}}'> \hat {{\text{SE}}}$
}
{
Set $\hat {{\text{SE}}}=\hat {{\text{SE}}}'$\;
}
{
Switch ${{\bf u}_{j,{\text{opt}}}}$ to the previous path in ${{\mathbb U}_j}(t)$\;
}
}
Set $j=j+1$\;
}
Generate tracking codebooks in (\ref{e30})\;
Achieve ${\hat{\bf f}_j}( t )$ and ${\hat{\bf w}_j}( t )$ via beam tracking\;
\KwOut{Optimal beam pair ${\hat{\bf f}_j}( t )$ and ${\hat{\bf w}_j}( t )$\;}
\end{algorithm}

The beam selection process unfolds as follows. Given that the beamforming vectors ${\bf f}_j (t)$ and ${\bf w}_j (t)$ in (\ref{e5}) are oriented towards particular directions, it becomes essential for each UE to pinpoint the most pertinent prior information for a singular path from the available ${\hat {L}_j}$ paths.
Initiating the selection, we designate ${{\bf u}_{j,{\text{opt}}}}={{\bf u}_{j,1}}$ as the preliminary choice for each UE, inherently prioritizing the path with the most significant path gain. The beams selected in this context, denoted by ${{\overline {\bf f}}_j} (t)$ and ${{\overline {\bf w}}_j}(t) $, are directed towards $\lceil {{{\hat \theta'_{j,{\text{opt}}}}}/{\pi }} \rceil$ and $\lceil {{{\hat \phi ’_{j,{\text{opt}}}}}/{\pi }} \rceil$, respectively, given as
\begin{subequations}\label{e26}
\begin{align}
{{\overline{\bf f}}_j} {\left(t\right)}&= {\boldsymbol{\sf{f}}_{M_j\left\lceil {{{\hat \theta ’_{j,{\text{opt}}}}}/{\pi }} \right\rceil }},\\
{{\overline {\bf w}}_j}{\left(t\right)} &= {\boldsymbol{\sf{w}}_{M_j\left\lceil {{{\hat \phi ’_{j,{\text{opt}}}}}/{\pi }} \right\rceil }},
\end{align}
\end{subequations}
where the expected codebook angular resolution $\pi/M_j=\pi/\left\lceil {{{M\pi }}/{{{{{S}}}_{j,{\text{opt}}}}}} \right\rceil $ is extended based on the angle searching range ${{{{S}}}_{j,l}}$. {\color{black}Thus, the predicted SE can be calculated as\footnote{{\color{black}The predicted SE serves as the optimization target in the beam management problem. Thus, the predicted SE is based on channel estimation in (24).}}
\begin{equation}\label{e27}
\hat{{\text{SE}}} = \mathop \sum \limits_{j = 1}^J {\text{lo}}{{\text{g}}_2}{{\left( {1 + \frac{{{{{\left| {\overline {\bf w}_j^{\text{H}}(t){{{\hat{\bf H}}}_j}(t){{\overline {\bf f}}_j}(t)} \right|}}^2}}}{{\mathop \sum \nolimits_{i \ne j} {{{\left| {\overline {\bf w}_j^{\text{H}}(t){{{\hat{\bf H}}}_j}(t){{\overline {\bf f}}_i}(t)} \right|}}^2} + {\sigma ^2}}}} \right)}}.
\end{equation}}

The analysis of downlink SE in this study reveals a critical limitation: the selection of paths with closely aligned values of ${\hat \theta'_{j,{\text{opt}}}}$, which may precipitate significant IUI. To mitigate this, an iterative refinement process for the selected prior information is executed across all $J$ UEs.
For each UE, identified as the $j$-th UE, the process involves the computation of the minimum radian distance, $\theta_{{\text{IUI}},j}$, to gauge the interference contribution of this particular UE. This computation is articulated as
\begin{equation}\label{e28}
\theta_{{\text{ IUI}},j}=\mathop {{\text{min}}}\limits_{i=1,\ldots,J,\,i \ne j} {\left| \hat {\theta}’ _{j,{\text{opt}}} - {\hat {\theta}’ _{i,{\text{opt}}} } \right|}.
\end{equation}
Should $\theta_{{\text{IUI}},j}<\theta _{\text{th}}$, the designated threshold for acceptable AoD radian distance, the $j$-th UE's IUI is deemed excessive. Consequently, an attempt is made to alternate ${{\bf u}_{j,{\text{opt}}}}$ to the subsequent path within ${{\mathbb U}_j}(t)$. The viability of this shift is evaluated by calculating the expected SE post-alteration, $\hat{{\text{SE}}}'$, employing (\ref{e26}) and (\ref{e27}).
The alteration is validated if $\hat{{\text{SE}}}'> \hat{{\text{SE}}}$. Otherwise, the initial selection remains unchanged. Through this iterative procedure, every UE attains optimal path information ${\hat \theta'_{j,{\text{opt}}}}, ~j=1, \ldots, J$, effectively reducing the potential for IUI and optimizing the overall SE of the system.

Then, the beam tracking operation, leveraging prior information, commences. According to the codebooks referenced in (\ref{e5}), $M^2$ beam measurements enable comprehensive beam sweeping across the angle region $(0,\pi)$. The beam tracking operation is designed to utilize the same overhead resources as traditional beam sweeping, facilitating the use of codebooks with superior angular resolution compared to $\pi/M$ and focusing on specific angular regions. Consider the case of the $j$-th UE. Initially, the angular resolution of the tracking codebooks, $\pi/{M_{{\text{tra}}}^j}$, is determined as
\begin{equation}
M_{{\text{tra}}}^j = {2^{{m_j}}}M,
\end{equation}
where ${m_j} = \mathop {{\text{arg\;min}}}\limits_{m\in \mathbb{N} } {\left| {{{{S}}}_j - \pi /{2^m}} \right|}$.
The tracking codebooks are designed to cover the angular region centered on ${\overline {\theta '} _j} = {{M_{{\text{tra}}}^j \lceil{{\hat \theta }'}_{j,{\text{opt}}}}}/{\pi }\rceil$ and ${\overline {\phi '} _{j}} = {{M_{{\text{tra}}}^j\lceil{{\hat \phi }’}_{j,{\text{opt}}}}}/{\pi }\rceil$, encompassing $\pi /{2^m}$. Therefore, the tracking codebooks are defined as
\begin{subequations}\label{e30}
\begin{align}
{{\bf f}_j}(t) \in& \bigg\{ { {\boldsymbol{\sf{f}}_{m + {\overline {\theta '} _j} }}{\bigg{|}}m = - \frac{M}{2} + 1, \ldots ,\frac{M}{2}} \bigg\},\\
{{\bf w}_j}(t) \in& \bigg\{ { {\boldsymbol{\sf{w}}_{m + {\overline {\phi '} _j} }}{\bigg{|}}m = - \frac{M}{2} + 1, \ldots ,\frac{M}{2}} \bigg\}.
\end{align}
\end{subequations}
where the beamforming vectors ${{\boldsymbol{\sf{f}}_{m+ {\overline {\theta '} _j}}}\in {\mathbb{C}^{ {N_{{\text{BS}}}} \times{1} } }}$ and ${\boldsymbol{\sf{w}}_{m+ {\overline {\phi '} _j}}}\in {\mathbb{C}^{ {N_{{\text{UE}}}} \times{1} } }$ are oriented in the $\frac{m+ {\overline {\theta '} _j}}{M}\pi$ and $\frac{m+ {\overline {\phi '} _j}}{M}\pi$ directions, featuring a codebook angular resolution of $\pi/{M_{{\text{tra}}}^j}$.
Upon completion of beam tracking for the $J$ UEs, the beam pair yielding the highest beamforming gain is selected as ${\hat{\bf f}_j}( t )$ and ${\hat{\bf w}_j}( t )$. This process finalizes the design for sensing-aided beam management, detailed in Algorithm \ref{A4}.


\section{Simulation Results}
This section details the performance evaluation of the proposed cooperative multi-user SLAM and beam management algorithm, emphasizing its application in radio map construction, UE localization, and beam management. The algorithm's efficiency is assessed in both indoor and outdoor settings, where mmWave signal reflections are universal, and light reflections can be omitted \cite{a2,R4C601}.
The indoor environment simulates a confined space, such as a room, where UEs represent pedestrians moving in random directions and steps. Conversely, the outdoor environment models a street scenario, with UEs depicting vehicles traveling in defined lanes at varying speeds. To substantiate the algorithm's performance, 1,000 Monte Carlo simulations were executed across these environments, considering variations in the number of UEs, FoVs, and feature dynamics (birth or death).

The evaluation metrics include: The mean Optimal Sub-Pattern Assignment (OSPA) in \cite{b0,a2,a4}, based on the Euclidean metric and using a cutoff parameter $c_{\text{map}}=10\;{\text{m}}$ and order 1, is used to assess the accuracy of radio map construction, given as 
\begin{equation}\label{R2C9_1}\small
\begin{aligned}
{\text{OSPA}} &(t)=\\
&\frac{1}{L_{\max}}
\left(
\min_{\varepsilon \in {\rm{\Pi }}_L }
\left\{
\sum \limits_{l= 1}^{L_{\min}} 
d\left( \hat{\bf{x}}_{{\text{VA}},l} (t) ,{\bf{x}}_{{\text{VA}},\varepsilon(l)}  \right)
\right\}
+
c_{\text{map}}L_{\Delta}
\right),
\end{aligned}
\end{equation}
where $L_{\max}=\max\left\{ L, |\mathbb{Q}(t)| \right\}$, $L_{\min}=\min\left\{ L, |\mathbb{Q}(t)| \right\}$, and $L_{\Delta}=| L- |\mathbb{Q}(t)| |$. The distance function is defined as $d(\mathbf{a},\mathbf{b})=\min \left\{ \|\mathbf{a} -\mathbf{b}\|_2 ,c_{\text{map}} \right\}$. The set ${{\rm{\Pi }}_{L} }$ consists of all possible permutations ${\varepsilon}$ on $\{1,\ldots, L\}$. 
The mean average UE localization error, $e_{\text{UE}}(t)$, is used to gauge UE localization precision. The SE, calculated by (\ref{e6}), is used to measure the efficiency of multi-user beam management. Details on the simulation setup and a comprehensive analysis of the results are presented in the following subsections.


\begin{table}[]
\renewcommand{\arraystretch}{1.2}
\caption{Measurement Settings}\label{t3}
\begin{tabular}{lll}
\toprule
 Measurement &  Parameter& Setting Justification  \\ \midrule
 Channel path angle &  $\sigma_{\text{angle}} = 0.04\,{\text{rad}}$ &Experiments in \cite{a4}  \\
 UE accelerometer & $\sigma _{{\text{IMU}}}=0.02\;{\text{m}/{\text{s}}^2}$  & Experiments in \cite{a13} \\
 Stereo camera images  &Default setting&    Refer to CARLA \cite{CARLA}\\
\bottomrule
\end{tabular}
\end{table}

\begin{table}[]
\caption{Scenario Settings}\label{t2}
\begin{tabular}{clll}
\toprule
Index & \multicolumn{1}{c}{FoV of UE$_1$} & \multicolumn{1}{c}{FoV of UE$_2$} & \multicolumn{1}{c}{FoV of UE$_3$} \\ \midrule
1 & PA, VA$_{2}$, VA$_{4}$ & PA, VA$_{1}$, VA$_{2}$, VA$_{4}$ & VA$_{1}$ - VA$_{4}$ \\
2 & PA, VA$_{2}$, VA$_{4}$ & \begin{tabular}[c]{@{}l@{}} VA$_{2}$ for \\$t \in (0,60)\,{\text{s}}$\end{tabular} & VA$_{1}$ - VA$_{4}$ \\
3 & PA, VA$_{2}$, VA$_{4}$ & \begin{tabular}[c]{@{}l@{}} VA$_{2}$ for \\$t \in (60,150)\,{\text{s}}$\end{tabular} & VA$_{1}$ - VA$_{4}$ \\
4 & PA, VA$_{2}$, VA$_{4}$ & PA, VA$_{1}$, VA$_{2}$, VA$_{4}$ & \begin{tabular}[c]{@{}l@{}} VA$_{3}$ for \\$t \in (0,60)\,{\text{s}}$\end{tabular} \\
5 & PA, VA$_{2}$, VA$_{4}$ & PA, VA$_{1}$, VA$_{2}$, VA$_{4}$ & \begin{tabular}[c]{@{}l@{}} VA$_{3}$ for \\$t \in (60,150)\,{\text{s}}$\end{tabular} \\
\bottomrule
\end{tabular}
\end{table}

\subsection{Indoor Scenario}
This subsection elaborates on a simulation set within a square room measuring $10\times10\,{\text{m}^2}$. The PA is centrally located within the room with $\mathbf{x}_{\text{PA}} = [0, 0]\,{\text{m}}$. Four VAs are generated by the room's walls as $\mathbf{x}_{\text{VA},1} = [-10,0]\,{\text{m}}$, $\mathbf{x}_{\text{VA},2} = [0,10]\,{\text{m}}$, $\mathbf{x}_{\text{VA},3} = [0,-10]\,{\text{m}}$, and $\mathbf{x}_{\text{VA},4} = [10,0]\,{\text{m}}$.

\subsubsection{Simulation Settings}
We simulate the movement of $J=3$ UEs, each taking steps of lengths between $0.3$ and $0.5$ meters in random directions over $T=150\,{\text{s}}$. The UEs' initial location estimates have an error of $e_\text{UE}(0)=0.15 \,{\text{m}}$, while the PA's location is precisely known, serving as the algorithm's starting point. 
The measurement settings are given in TABLE \ref{t3}. The channel path angle estimation noise satisfies $\sigma_{\text{angle}} = 0.04\,{\text{rad}}$ according to experiments in \cite{a4}. The IMU measurements include the UE accelerometer data in \cite{a13} with noise variance $\sigma _{{\text{IMU}}}=0.02\;{\text{m}/{\text{s}}^2}$. The vision measurements are generated based on the stereo camera in CARLA \cite{CARLA} with default settings. 
Algorithm parameters consist of the stage switch time $T_{\text{config}}=10\,{\text{s}}$, allowing sufficient time for UEs to achieve preliminary local SLAM convergence. Additionally, the data association distance threshold $r_{\text{cut}}=0.3$ is employed, scaling with the scenario size, and the fixed pruning threshold $r_{\text{cut}}=0.3$ is set to mitigate false alarms and missed detections. Due to privacy concerns, stereo cameras are not deployed in this scenario.

To assess the algorithm's accuracy in constructing radio maps, five scenarios with varying UEs' FoVs and feature dynamics (birth/death) are introduced, detailed in TABLE \ref{t2}.
Scenario 1 establishes a baseline with UEs possessing distinct FoVs and no features undergoing birth or death processes. Scenarios 2 through 5 build upon this foundation, introducing variations in feature dynamics:
\begin{itemize}
\item In Scenario 2, UE$_{2}$ is limited to observing VA$_{2}$ only within the first 60 seconds. The PA and VA$_{1}$ and VA$_{4}$ remain consistently observable. This configuration results in the ``death'' of a local feature for UE$_{2}$ as it surpasses the 60-second mark.
\item Scenario 3 explores a similar setup as Scenario 2, but it focuses on the ``birth'' of a local feature.
\item Scenario 4 examines the ``death'' of a global feature from the BS's perspective, induced by UE$_{3}$'s exclusive visibility of VA$_{3}$ ceasing after 60 seconds, whereas VA$_{1}$, VA$_{2}$, and VA$_{4}$ remain visible throughout.
\item Conversely, Scenario 5 investigates the ``birth'' of a global feature under analogous circumstances to Scenario 4.
\end{itemize}
These scenarios collectively assess the algorithm's adaptability to different UEs observing non-overlapping features, as well as its responsiveness to the birth and death of both local and global features.


\begin{figure}
\centering
\subfigure[]{
\hspace{-10mm} \includegraphics[scale=0.37]{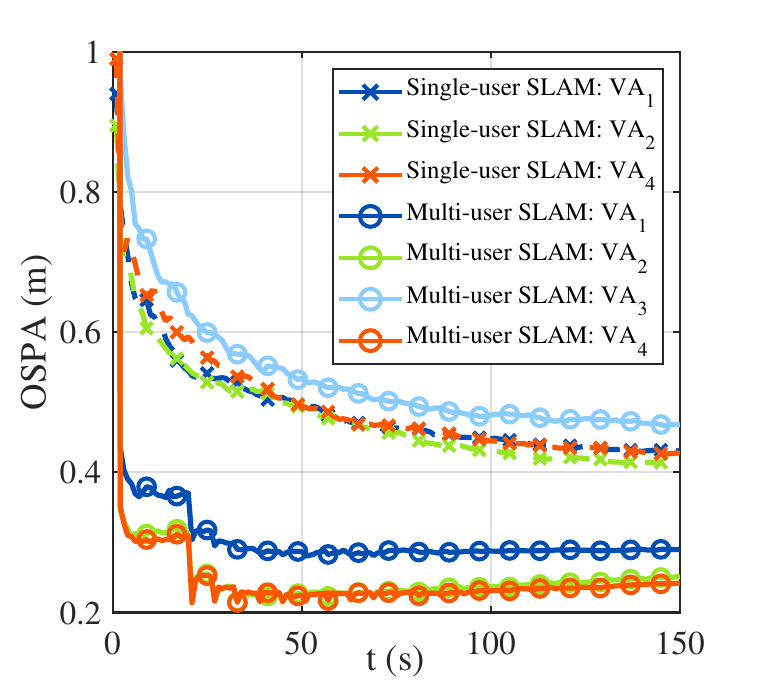}\hspace{-6mm}
}
\quad
\subfigure[]{
\hspace{-6mm} \includegraphics[scale=0.37]{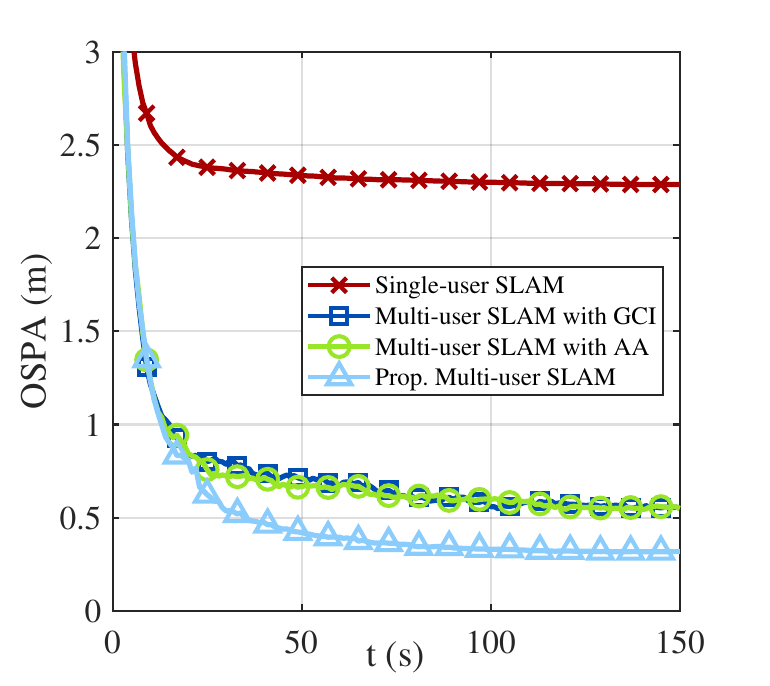}\hspace{-12mm}
}
\caption{Performance of radio map construction: (a) across different features and (b) compared with various algorithms in scenario 1.} 
\label{f4}\vspace{-0.3cm}
\end{figure}

\begin{figure}
\centering
\subfigure[]{
\hspace{-10mm} \includegraphics[scale=0.37]{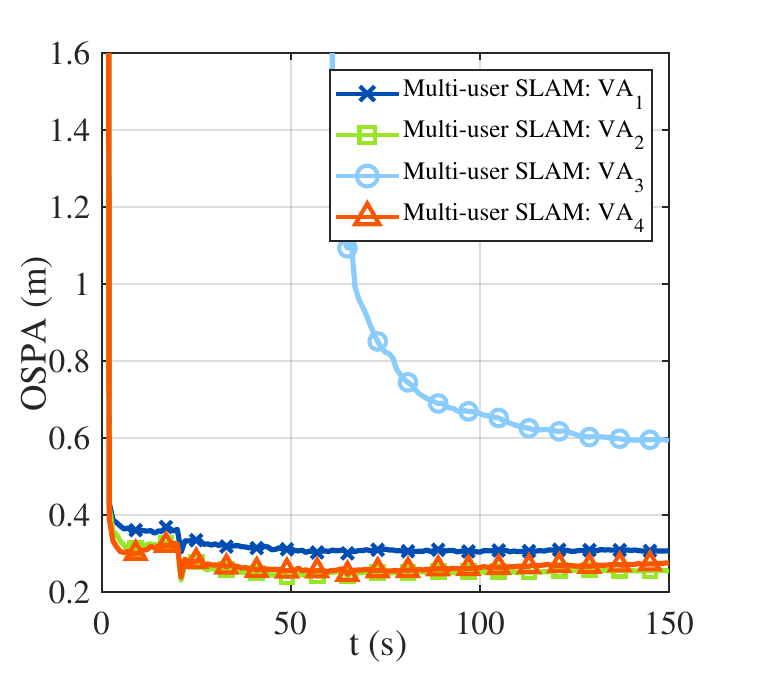}\hspace{-6mm}
}
\quad
\subfigure[]{
\hspace{-6mm} \includegraphics[scale=0.37]{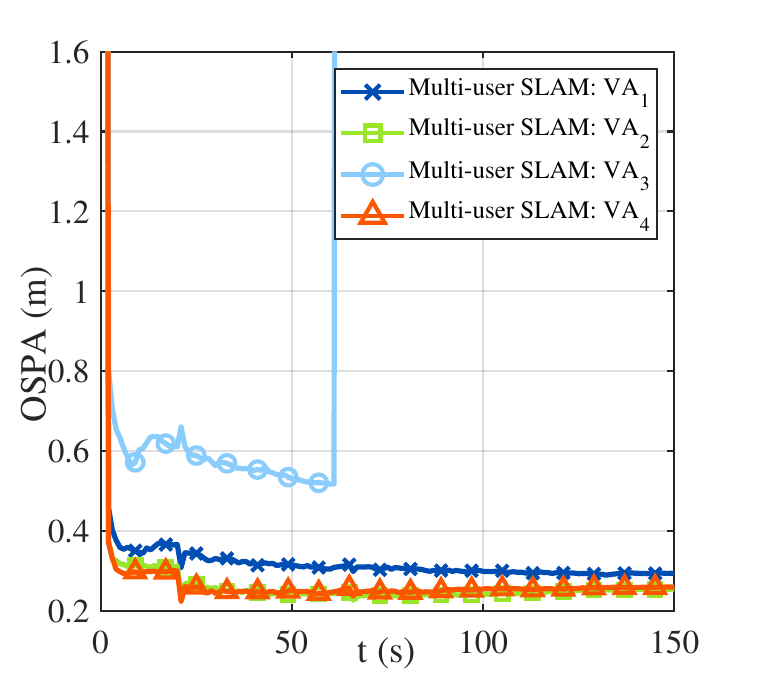}\hspace{-12mm}
}
\caption{Performance of radio map construction in (a) scenario 4, which considers global feature birth, and (b) scenario 5, which considers global feature death.}
\label{f5}\vspace{-0.3cm}
\end{figure}

\subsubsection{Analysis and Discussion}
\begin{figure}
\centering
\subfigure[]{
\hspace{-10mm} \includegraphics[scale=0.38]{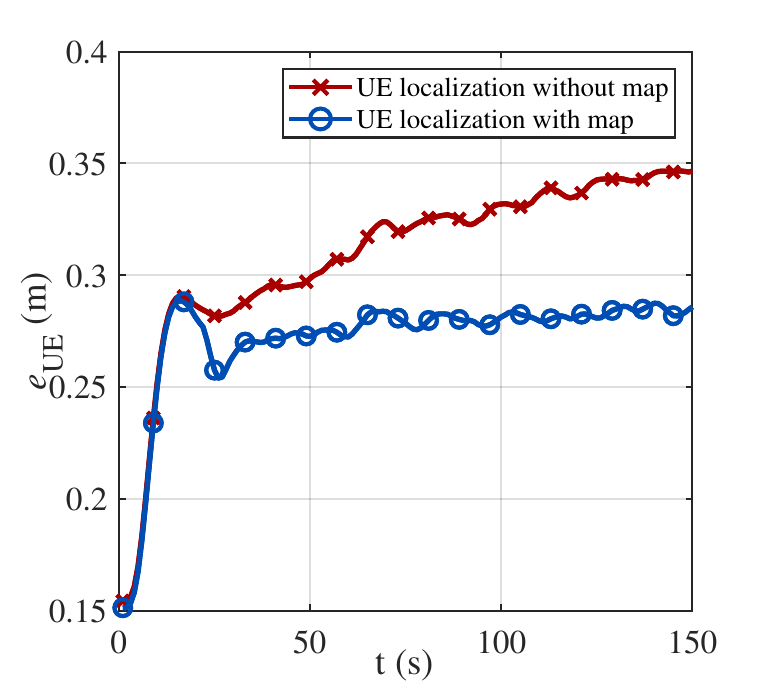}\hspace{-3mm}
}
\quad
\subfigure[]{
\hspace{-9mm} \includegraphics[scale=0.38]{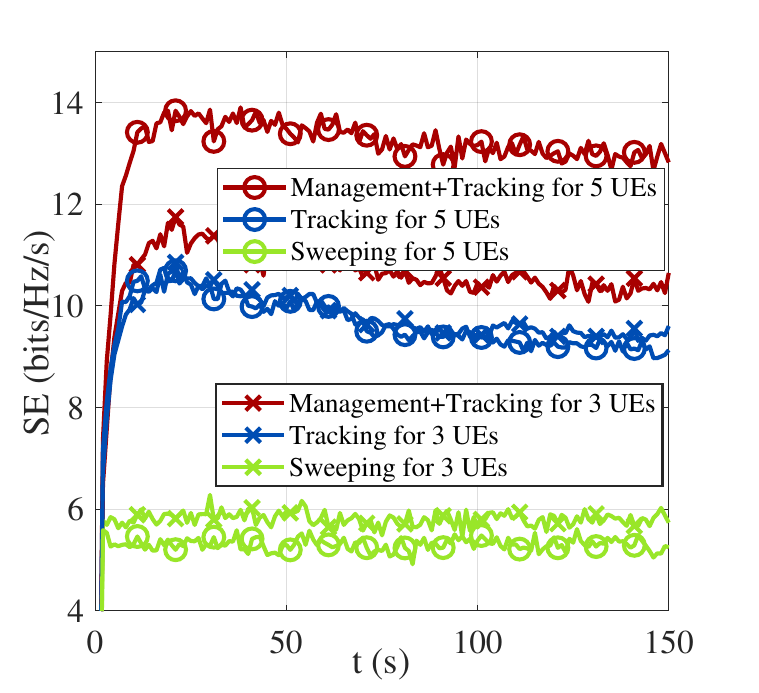}\hspace{-10mm}
}
\caption{(a) Average UE localization performance among different feature birth or death scenarios. (b) SE curves of different beam management algorithms in scenarios of 3 UEs and 5 UEs.}
\label{f6}\vspace{-0.4cm}
\end{figure}

Curves of $e_{\text{UE}}(t)$, ${\text{OSPA}}(t)$, and ${\text{SE}}(t)$ are drawn in different scenarios, as illustrated in Figs. \ref{f4}, \ref{f5}, and \ref{f6}.
The radio map construction performance, UE localization accuracy, and beam management efficiency are evaluated as follows.

We first compared the proposed multi-user SLAM algorithm, executed by all UEs, with the SLAM algorithm performed solely by UE$_{2}$ in scenario 1, to assess the accuracy of global radio map construction.
The OSPA curves for the specific features are depicted in Fig. \ref{f4}(a).
For the single-user case, represented by the dotted lines labeled ``Single-user SLAM,'' only a portion of the features can be estimated due to the limited FoV of UE$_{2}$. The average map construction accuracy for UE$_{2}$ is $0.52\,{\text{m}}$.
For the multi-user case, represented by the solid lines labeled ``Multi-user SLAM,'' all VAs can be estimated, as the combined FoVs of all three UEs cover all VAs. The radio map in the multi-user case achieves an average accuracy of $0.27\,{\text{m}}$ for estimating VA$_{1}$, VA$_{2}$, and VA$_{4}$, which is $48.5\%$ higher than that in the single-user case.
Additionally, the accuracy of VA$_{3}$ estimated in the multi-user case is similar to that in the single-user case. This similarity is because only UE$_{3}$ can observe VA$_{3}$, meaning no integration is possible for VA$_{3}$. Consequently, the accuracy of VA$_{3}$'s estimation relies solely on UE$_{3}$'s observations.

The average OSPA curves for all available features in scenario 1 are depicted in Fig. \ref{f4}(b), where we compared the map construction accuracy of the proposed multi-user SLAM algorithm with two existing methods: the AA and GCI integration methods \cite{a6}.
The single-user case, illustrated in red in Fig. \ref{f4}(b), can only observe a subset of features, resulting in an incomplete radio map construction.
Moreover, the proposed algorithm outperforms both the AA and GCI integration methods, showing significant improvement in multi-user scenarios. This improvement stems from the two-stage design of the proposed algorithm, which effectively combines the strengths of both improved AA and GCI integration techniques.
We implemented the proposed algorithm in scenarios 2-5 to evaluate the global radio map’s adaptability to the birth or death of VAs.
In scenarios 2 and 3, the birth or death of local feature VA$_2$ in UE$_2$ is considered, respectively. However, the proposed algorithm still maintains comparable radio map construction performance to scenario 1{\footnote{The OSPA curves in scenarios 2-3 are omitted due to space limits. These OSPA curves are similar to the ``Prop. Multi-user SLAM'' curve in Fig. \ref{f4}(b).}}.
This is because other UEs can still observe VA$_2$, and the corresponding integration functions without UE$_{2}$'s observations.
In scenarios 4 and 5, the birth and death of global feature VA$_3$ occur. The OSPA curves for scenarios 4 and 5, shown in Figs. \ref{f5}(a) and \ref{f5}(b), respectively, confirm the algorithm's adaptability to the birth and death of global features.
Therefore, the proposed algorithm can effectively manage the birth and death of both local and global features.

We then consider different FoV conditions for UEs to evaluate the UE localization performance.
First, UE$_{1}$ and UE$_{3}$ in scenario 1 were examined, representing FoV conditions with limited and sufficient VA observations, respectively.
Next, UE$_{2}$ in scenarios 2 and 3 was assessed, representing FoV conditions with the birth and death of local features.
Finally, UE$_{3}$ in scenarios 4 and 5 was considered, representing FoV conditions with the birth and death of global features.
The average $e_{\text{UE}}(t)$ curves of these scenarios are illustrated in Fig. \ref{f6}(a).
For UEs participating in the multi-user SLAM algorithm, indicated as ``UE localization with map'' in red lines in Fig. \ref{f6}(a), a localization error of $0.28\,{\text{m}}$ is achieved. This represents a $15.7\%$ improvement over the $0.33\,{\text{m}}$ localization error obtained by UEs without access to the global radio map, indicated as ``UE localization without map'' in blue lines in Fig. \ref{f6}(a).
These results demonstrate that the proposed multi-user SLAM algorithm can improve the UE localization accuracy under various FoV conditions.

Finally, we consider a specific scenario where $J$ UEs can observe all four features with a changeable UE number $J$ to evaluate the beam management efficiency.
We set $J=3$ or $5$ and performed the proposed algorithm for analysis, with the codebook length $M=8$.
Three control groups consuming the same beam measurement number $M^2=64$ are considered.
The first one considers the proposed sensing-aided beam management algorithm based on (\ref{e30}), denoted as ``Management+Tracking''.
The second one considers the beam management strategy based on ${{\bf u}_{j,{\text{opt}}}}={{\bf u}_{j,1}}$ \cite{a13}, which means that all UEs select their local optimal beam pairs. This control group is denoted as ``Tracking''.
The third one considers that each UE selects the most powerful beam pair through beam sweeping based on (\ref{e5}), denoted as ``Sweeping''.
Fig. \ref{f6}(b) illustrates the ${\text {SE}}(t)$ curves in scenarios of 3 UEs and 5 UEs. The red curves representing ``Management+Tracking'' can achieve a higher SE compared with ``Tracking'' and ``Sweeping'' respectively in blue and green, especially in the scenario of 5 UEs. The SE of ``Management+Tracking'' can be improved by $149\%$ compared with ``Sweeping'', and $36\%$ compared with ``Tracking'' for 5 UEs.
Moreover, there is no significant change in the SE curves of ``Tracking'' and ``Sweeping'' when the UE number increases.
This is because the increased UE density raises the likelihood of overlapping downlink beams at the BS, leading to significant IUI.
However, the SE of ``Management+Tracking'' can be improved with the increase of the UE number, which proves the anti-IUI ability of the proposed prior information selection process.


\subsection{Outdoor Scenario}
\begin{figure}
\centering
\subfigure[]{
\hspace{-2mm} \includegraphics[scale=0.43]{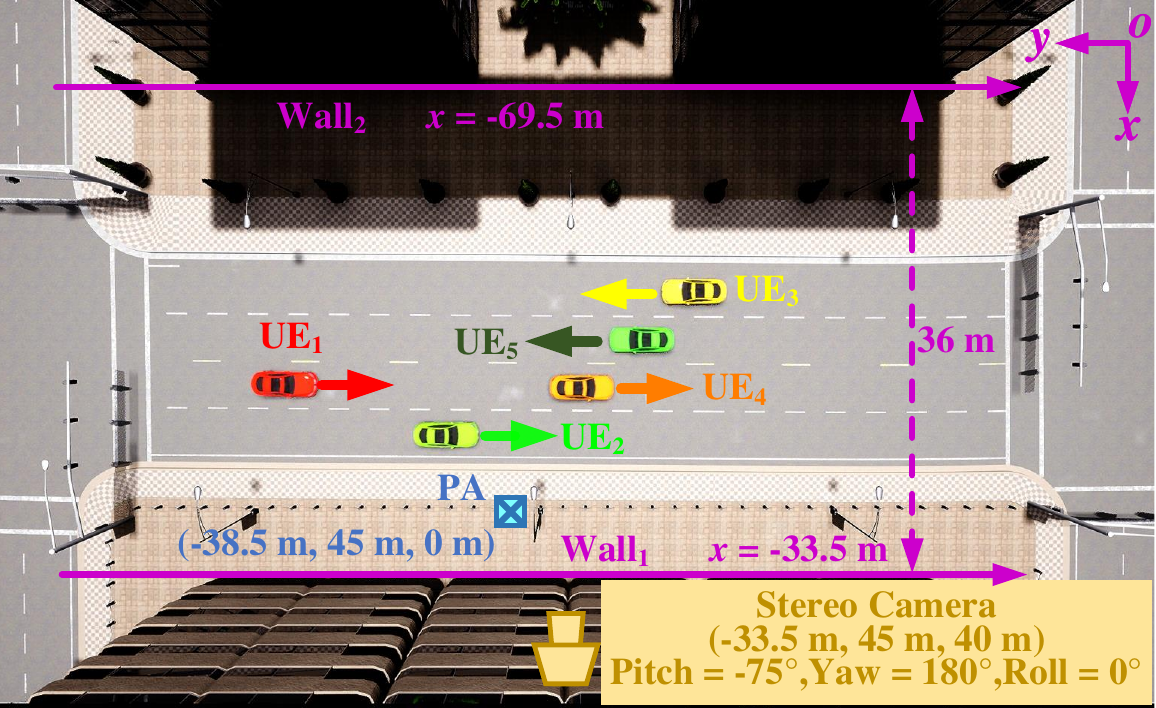}\hspace{0mm}
}
\quad
\subfigure[]{
\hspace{-2mm} \includegraphics[scale=0.185]{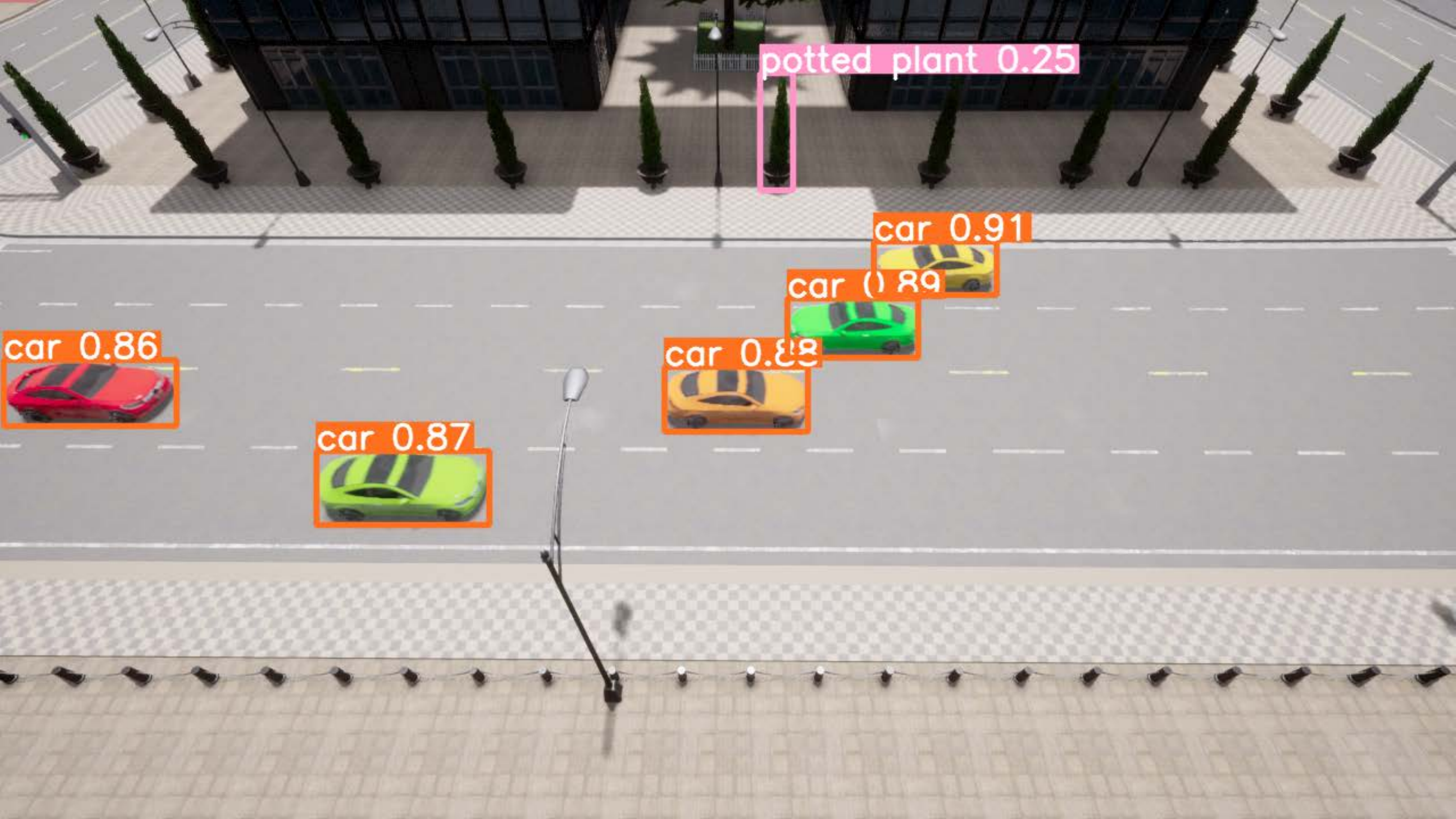}\hspace{0mm}
}
\caption{(a) the CARLA simulation settings and (b) an illustration of YOLO detections.}
\label{f8}\vspace{-0.3cm}
\end{figure}

This section considers an outdoor street with a road width of $34\;{\text{m}}$, constructed in CARLA \cite{CARLA}, as illustrated in Fig. \ref{f8}(a). UEs refer to vehicles on specific lanes in the outdoor environment. The PA and UEs are positioned close to the ground, and thus, only azimuth angles are considered.
The PA, shown in blue, is positioned at the sidewalk of the street with $\mathbf{x}_{\text{PA}} = [-38.5, 45]\,{\text{m}}$.
The two buildings along the street are assumed to provide two flat walls, as illustrated in purple. Thus, two VAs can be generated at $\mathbf{x}_{\text{VA},1} = [-100.5, 45]\,{\text{m}}$ and $\mathbf{x}_{\text{VA},2} = [-28.5, 45]\,{\text{m}}$.

\begin{table}[]
\renewcommand{\arraystretch}{1.2}
\caption{Differences in Simulation Settings\\ between Indoor and Outdoor Scenarios}\label{t4}
\begin{tabular}{lll}
\toprule
Simulation Settings & Indoor Scenario   & Outdoor Scenario \\
\midrule
UE velocity & $[0.3,0.8]$ m/s & $[30,50]$ km/h \\
Initial localization error & 0.15 m & 0.8 m \\
Simulation site size & $10 \times 10$ m$^2$ & $36 \times 55$ m$^2$ \\
Number of anchors & 4 & 2 \\
Anchor locations & \begin{tabular}[c]{@{}c@{}} $[-10,0]$ m,\ $[0,10]$ m \\ $[0,-10]$ m,\ $[10,0]$ m \end{tabular} & \begin{tabular}[c]{@{}c@{}} $[-100.5,45]$ m \\ $[-28.5,45]$ m \end{tabular} \\
Device configuration & \begin{tabular}[c]{@{}c@{}} mmWave array, \\ IMU \end{tabular} & \begin{tabular}[c]{@{}c@{}} mmWave array, \\ IMU, stereo camera \end{tabular} \\
\bottomrule
\end{tabular}
\end{table}

\subsubsection{Simulation Settings}
A total of $J=5$ UEs with different colors are positioned in this section. These UEs move at random speeds across four different lanes on the street. Their moving time varies across different Monte-Carlo experiments, and the shortest time is selected for analysis. 
The main simulation setting differences between indoor and outdoor scenarios are given in TABLE \ref{t4}. The outdoor UEs are assumed to be communication devices mounted on vehicles, with the vehicle speed ranging $[30,50]\;{\text{km/h}}$ and an initial localization average error of $0.8\;{\text{m}}$. Compared with indoor UEs, outdoor UEs with higher speeds and larger initial errors pose greater challenges for UE localization. The outdoor simulation site covers an area of $36\times55$ m$^2$, with a total of two VAs. Compared with the indoor site, the outdoor site is larger with sparse VAs, making radio map construction more challenging. Moreover, the outdoor scenario is additionally equipped with a fixed stereo camera enabling multi-modal enhanced UE localization. 
 
The stereo camera is positioned at the top of the building with a known location and orientation, as illustrated in yellow in Fig. \ref{f8}(a). The camera's internal settings are the default in CARLA, and its RGB image is shown in Fig. \ref{f8}(b). 
We applied a pre-trained YOLOv8 model \cite{YOLOv8} for object detection, and the detected cars are used for camera-based multi-object localization. The sensing performance metrics ${\text{OSPA}}(t)$ and ${e_\text{UE}}(t)$ are considered in this section. The radio map construction and UE localization performance of the proposed algorithm are evaluated, while the beam management analysis is omitted due to similar results.

\begin{figure}
\centering
\includegraphics[scale=0.55]{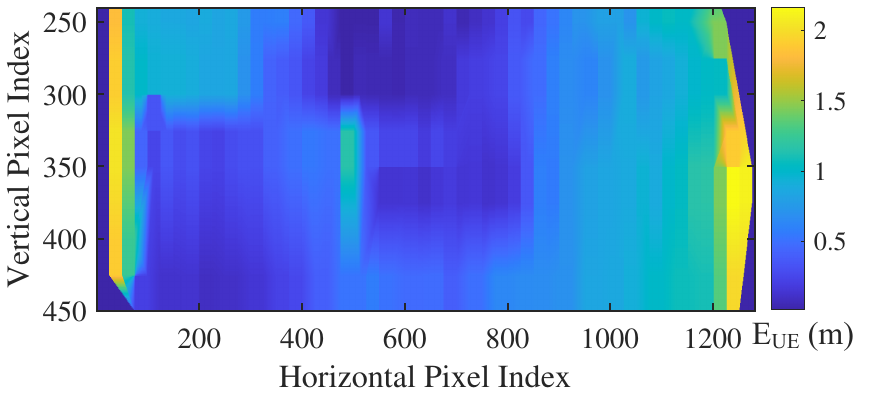}
\caption{Visualization of the look-up table that maps the input pixel coordinates with the localization reliability.}
\label{f9}\vspace{-0.4cm}
\end{figure}

\subsubsection{Analysis and Discussion}
In this subsection, we complete the offline training before performing the cooperative multi-user SLAM algorithm to enable the multi-modal enhanced localization in Section V. Specifically, a total of 1,000 vehicles traveling through the street comprise the training set, with their true trajectories available. The look-up table in (\ref{e21}) is realized via the training set. The heat map describing the relationship between the localization error squared variance and pixel coordinates is illustrated in Fig. \ref{f9}. 
Moreover, a validation set consisting of 1,000 vehicles traveling through the street is used to verify the effectiveness of the localization error model, yielding a mean deviation of $0.036\;{\text{m}}$. Camera-based multi-object localization achieves a minimum error of $0.005\;{\text{m}}$ at the center of the image and a maximum error of $2.0\;{\text{m}}$ at the image edge.
Achieving continuous and stable object localization in this manner is challenging. Furthermore, the camera cannot distinguish between communication UEs, making camera-based multi-object localization unsuitable for multi-user localization requirements. 

\begin{figure}
\centering
\subfigure[]{
\hspace{-12mm} \includegraphics[scale=0.38]{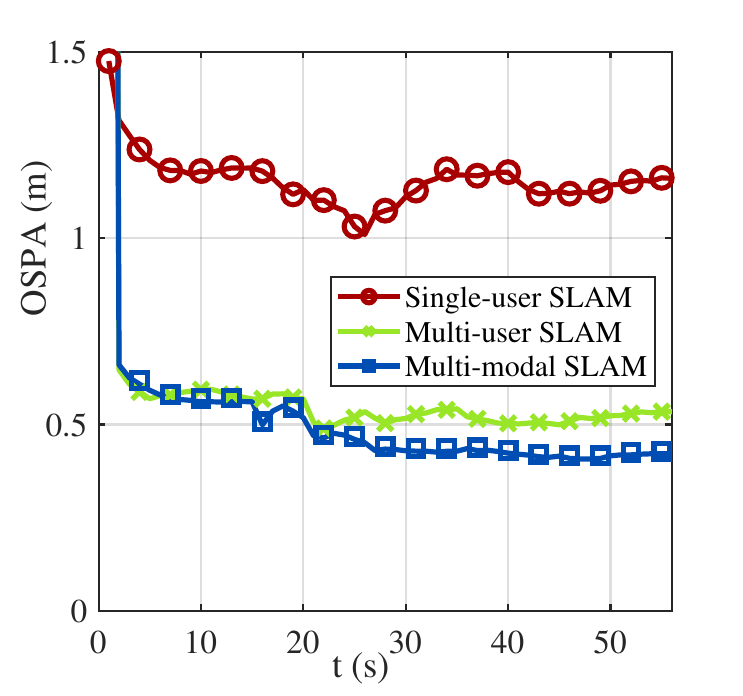}\hspace{-6mm}
}
\quad
\subfigure[]{
\hspace{-5mm} \includegraphics[scale=0.38]{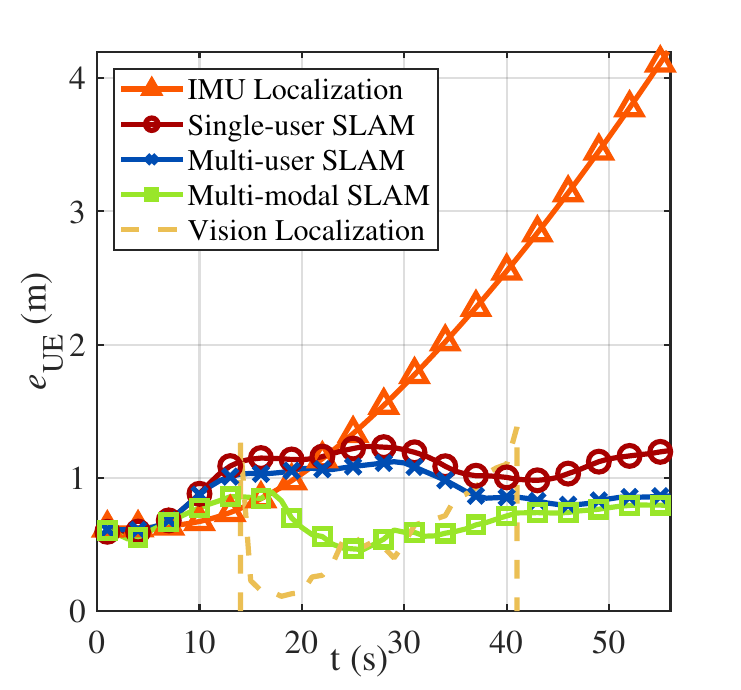}\hspace{-12mm}
}
\caption{Performance comparison for different SLAM algorithms: (a) radio map construction performance, and (b) UE localization performance.}
\label{f10}\vspace{-0.5cm}
\end{figure}

Utilizing the look-up table generated from the offline training process, the multi-modal SLAM can be implemented. A total of $J=5$ UEs observing both VA$_1$ and VA$_2$ is considered.
We compare the proposed multi-modal SLAM algorithm with the SLAM algorithm performed by a single UE, the multi-user SLAM algorithm considering only the radio-IMU measurements, and the localization results relying on IMU or vision measurements to evaluate the radio map construction and UE localization performance.
The OSPA curves are displayed in Fig. \ref{f10}(a). The curve in red represents the SLAM performed by a single UE, the curve in blue represents the proposed multi-user SLAM, and the curve in green represents the multi-modal SLAM.
The multi-modal SLAM attains an average OSPA of $0.44\;{\text{m}}$, demonstrating a marginal improvement over the multi-user SLAM. However, compared to single-user SLAM with an average OSPA of $1.13\;{\text{m}}$, the multi-modal SLAM achieves an improvement of $60.9\%$.

The UE localization performance is evaluated through the $e_{\text{UE}}(t)$ curves in Fig. \ref{f10}(b).
The orange curve represents the localization results relying only on IMU, the red curve represents the SLAM performed by a single UE, the blue curve represents the multi-user SLAM, the green curve represents the proposed multi-modal SLAM algorithm, and the brown curve represents localization results relying only on the stereo camera measurements.
For IMU localization results, the accumulated error would continuously increase $e_{\text{UE}}(t)$.
For camera localization results, the $e_{\text{UE}}(t)$ curve is not continuous due to the unexpected miss-detection of YOLO. These two sensors cannot offer reliable localization results alone.
The multi-modal enhanced localization outperforms the other two SLAM algorithms when $t\in (15,40)\,{\text{s}}$. This improvement is attributed to vehicles often reaching the middle of the street, where camera-based multi-object localization effectively complements SLAM localization results. The average localization accuracy of the multi-modal SLAM during this period is $0.69\;{\text{m}}$, which is an improvement of $27.1\%$ compared to the multi-user SLAM (with an error of $0.95\;{\text{m}}$), and $37.5\%$ compared to the SLAM performed by a single UE (with an error of $1.11\;{\text{m}}$).

\section{Conclusion}
This study presented a cooperative multi-modal SLAM framework, aimed at improved mapping, localization, and beam management. The proposed approach included a Bayesian framework for multi-user SLAM, supported by a two-stage algorithm for constructing a global radio map. Additionally, a multi-modal enhanced localization algorithm integrating radio, vision, and IMU measurements was developed. Using the global radio map and UE localization data, we introduced sensing-aided beam management to improve overall system efficiency.
Numerical analyses confirmed that the proposed algorithms outperformed traditional methods. Radio map construction accuracy improved by up to 60\% in comparison to single-user scenarios, while the multi-modal enhanced localization algorithm achieved a 37.5\% increase in accuracy over single-modal approaches. Sensing-aided beam management further demonstrated a 36\% improvement in SE over conventional methods. Notably, the proposed algorithms proved adaptable to both indoor and outdoor environments under various UE FoV conditions.

\bibliographystyle{IEEEtran}
\bibliography{multiSLAML_V2}

\end{document}